\documentclass[sn-mathphys,Numbered,referee]{sn-jnl}
\usepackage{lineno}
\pdfoutput=1 

\usepackage{graphicx}%
\usepackage{multirow}%
\usepackage{amsmath,amssymb,amsfonts}%
\usepackage{amsthm}%
\usepackage{mathrsfs}%
\usepackage[title]{appendix}%
\usepackage{xcolor}%
\usepackage{textcomp}%
\usepackage{manyfoot}%
\usepackage{booktabs}%
\usepackage{algorithm}%
\usepackage{algorithmicx}%
\usepackage{algpseudocode}%
\usepackage{listings}%
\usepackage{lscape}
\usepackage{multicol}

\usepackage{pifont} 
\usepackage{float} 

\theoremstyle{thmstyleone}%
\theoremstyle{thmstyletwo}%

\theoremstyle{thmstylethree}%

\begin{document}

\title[Article Title]{Detection of anomalous spatio-temporal patterns of app traffic in response to catastrophic events}


\author[1]{Sofia Medina}
\author[1]{Shazia'Ayn Babul}
\author[1]{Rohit Sahasrabuddhe}
\author[1]{Timothy LaRock}
\author[1]{Renaud Lambiotte}
\author[1]{Nicola Pedreschi}

\affil[1]{\orgdiv{Mathematical Institute}, \orgname{University of Oxford}, \country{UK}}


\abstract{In this work, we uncover patterns of usage mobile phone applications and information spread in response to perturbations caused by unprecedented events. We focus on categorizing patterns of response in both space and time and tracking their relaxation over time. To this end, we use the NetMob2023 Data Challenge dataset, which provides mobile phone applications traffic volume data for several cities in France at a spatial resolution of 100$m^2$ and a time resolution of 15 minutes for a time period ranging from March to May 2019. We analyze the spread of information before, during, and after the catastrophic Notre-Dame fire on April 15th and a bombing that took place in the city centre of Lyon on May 24th using volume of data uploaded and downloaded to different mobile applications as a proxy of information transfer dynamics. We identify different clusters of information transfer dynamics in response to the Notre-Dame fire within the city of Paris as well as in other major French cities. We find a clear pattern of significantly above-baseline usage of the application Twitter (currently known as X) in Paris that radially spreads from the area surrounding the Notre-Dame cathedral to the rest of the city. We detect a similar pattern in the city of Lyon in response to the bombing. Further, we present a null model of radial information spread and develop methods of tracking radial patterns over time. Overall, we illustrate novel analytical methods we devise, showing how they enable a new perspective on mobile phone user response to unplanned catastrophic events, giving insight into how information spreads during a catastrophe in both time and space.}

\keywords{Data visualization, Mobile applications, Event detection, Disaster response, Urban Mobility, Spatiotemporal phenomena, Patterns in Data
}



\maketitle

\section{Introduction}\label{sec1}

Understanding how information propagates during and after catastrophic events is an active field of investigation ~\cite{online_dynamics, novelty_collective_attention, lehmann_twitter_cluster,osborne2012bieber, he_measuring_2017}. Mobile phone data provides deep insight into the intricacies of human behavior, with extremely high granularity on both temporal and spatial scales. These data-sets enable large-scale data driven analysis applied to a wide range of areas including social network analysis \cite{eagle2009inferring}, population dynamics \cite{deville2014dynamic}, urban structure \cite{louail2014mobile}, public health \cite{oliver2020mobile}, and disaster response \cite{lu2012predictability}. Social media and online resources have been used to categorize \cite{lehmann_twitter_cluster} and track the duration and intensity \cite{osborne2012bieber, wikipedia_events, he_measuring_2017,Sune_accelerating} of responses to events and breaking news stories. Attention is also paid to the behavioral patterns that characterize and distinguish between planned and unplanned events ~\cite{kobayashi_modeling_2021, crane_robust_2008}. While most studies on social media or app usage have focused on a single type of online activity~\cite{lehmann_twitter_cluster,kobayashi_modeling_2021, crane_robust_2008,Jure_temporal}, recent work has emphasized the importance of cross-platform analysis to account for platform-specific effects~\cite{shen_information_2017,alipour_cross-platform_2024}.

In this work, we analyze mobile phone data to understand how the temporal and spatial usage of different applications are perturbed in the aftermath of an unprecedented event, and how these usage patterns evolve over time. We will focus on how a variety of mobile phone applications, which may have similar user-function, respond to the same single event. To this end, we use the NetMob2023 Data Challenge dataset \cite{netmob23} which provides mobile phone usage data for several cities in France for a range of applications over a period of time ranging from March to May of 2019 at a spatial resolution of 100$m^2$ and a time resolution of 15 minutes. We investigate dynamics and patterns arising from catastrophic events, most notably, one of the most extreme events occurring during this period: the fire at the Notre-Dame cathedral in Paris and the collapse of its historic spire.

The Notre-Dame cathedral, built during the 12th and 13th centuries, and has long been recognized as an emblem of French society. Located in the center of Paris, it is a highly popular tourist attraction, as well as a religious site. On April 15th, 2019, the roof of the Notre-Dame cathedral caught on fire and was severely burned. The fire broke out in the attic of the cathedral at 18:18 \cite{peltier_notre-dame_2019}.
The occupants of the building, including tourists and worshippers taking mass, evacuated minutes later, at 18:20 \cite{ap_news_66mins}, when a fire alarm sounded. Smoke was first visible by 18:43  from Paris' Left Bank \cite{ap_news_66mins}, with firefighters officially called to the cathedral at 18:51 \cite{bennhold_notre-dames_2019} when cathedral workers discovered the fire. The historic spire of the cathedral collapsed at 19:50 \cite{peltier_notre-dame_2019}. Given that the fire occurred at one of the most historic buildings in Europe, news spread quickly through both social media and traditional news media sources. 

We analyze the spread of information before, during, and after the Notre-Dame fire using the volume of data uploaded and downloaded to different mobile applications as a proxy for information transfer. The methods we develop can be extended to other contexts to characterize mobile phone user response to unplanned catastrophic events, giving insight into how information spreads during a catastrophe. We extend our spatial analysis to find consistent results in a second catastrophic event, a bombing in Lyon, France.

We present two kinds of analysis. First, we spatially aggregate data within each city and study the time-series of app traffic per city for a select subset of the available apps. We then use a simple anomaly detection method to determine whether traffic volume for each app spiked after the Notre-Dame fire, when those spikes occurred, and for how long. We use information about these spikes in traffic volume to cluster apps based on how they were used during the fire, attempting to discern patterns in type of application use both within the city of incidence of catastrophe and across other cities. We conclude that applications have a similar general user trend across cities, however there are variations in app traffic behavior that can be explained by the strong effect of city-specific user preferences. Additionally, we find that a fine-grained description of application function is a more accurate predictor of clustering behavior. Second, we analyze how traffic spikes were distributed spatially throughout the city of Paris over time as news of the fire spread and mobile phone users sought more information through social media, apps, and streaming services.
Most prior studies of information spread focus on the relationships between users in social-network space~\cite{newman2002spread, granovetter1978threshold, centola2010spread}.  There are few works, however, which investigate the geo-spatial dependence of spread of information over social media~\cite{yang2019reshaping}. We find that that the abnormally elevated traffic appears to spreads radially from the center of the catastrophe. We construct a null model to show data patterns shifts to radial data patterns in response to catastrophe and the relaxation of radial patterns over time. This is consistent with other literature~\cite{Bagrow_emergencies}, as well as our own extended analysis to a bombing occurring in Lyon, France. We propose methods for quantifying this radial pattern of the information spread, noting that it this pattern of information spread also extends to planned events.

\begin{figure}[H]
    \centering
    \includegraphics[width=\textwidth]{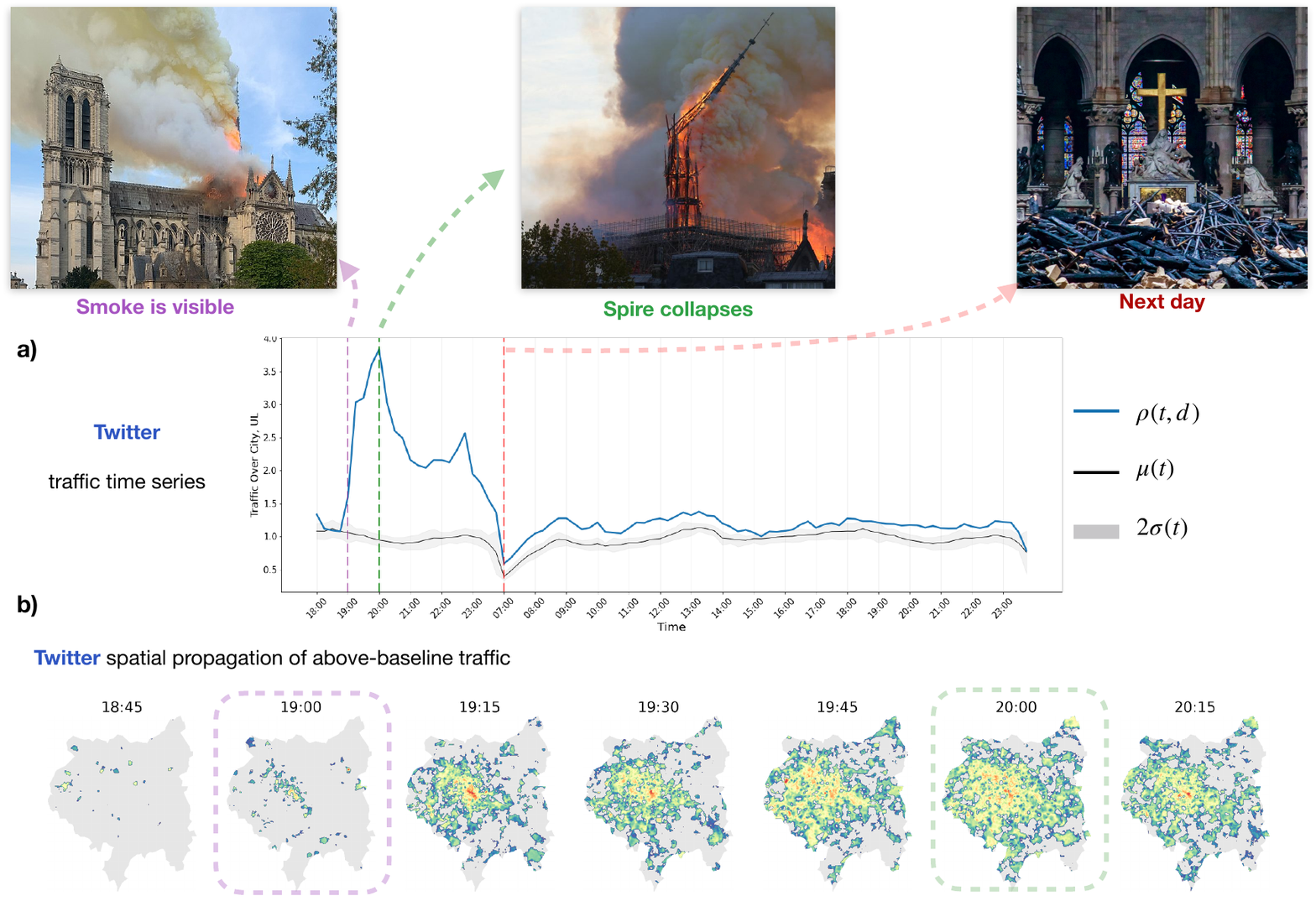}
    \caption{\textbf{Graphical introduction:} \textbf{a)} The time series evolution of the application Twitter from 18 : 00 to end of day on the day of the fire and 7 : 00 to end of day the day following the fire. The traffic on the day of the fire, $\rho$, is given in blue, and the mean ($\mu$) and two standard deviations ($2\sigma$) of traffic in previous weeks is in grey. \textbf{b)} The spatial evolution of application traffic of Twitter is shown in time on the outline of the city of Paris, with intensity of app traffic from an established baseline displayed.}
    \label{fig:Fig0}
\end{figure}

\section{Methods}\label{sec2}

In this section, we will describe the methodology pertaining to the detection and analysis of patterns of spikes in application traffic volume both temporally and spatially.

\subsection{Data}

The dataset under scrutiny is the NetMob2023 Data Challenge dataset \cite{netmob23}, which provides the traffic volume for 68 mobile applications in 20 cities in France at a spatial resolution of $100m^2$ and a time resolution of 15 minutes for 77 days, ranging from March 16th to May 31st 2019. The data contains both uplink (transmissions to the cell tower) and downlink (transmissions from the cell tower) traffic. We consider both of these traffic directions simultaneously by summing the two. In this work, we limit our analysis to data from Paris, Marseille, Lyon, Montpellier, Rennes, and Strasbourg. We note that the dataset does not contain information regarding content transferred along the cellular network. 

In our analysis, we consider a subset of applications in the dataset after sorting for relevance (e.g. exclusion of mobile phone games) and ability to distinguish app traffic as being in some form related to the fire (e.g. exclusion of general internet browsing such as Yahoo). A full list of applications considered, as well as excluded, is given in the Supplementary Materials.

\subsection{Detecting Spikes in App Traffic Volume}

In this dataset, each city $c$ is represented spatially by a tiling (or tessellation); we refer to the tiles of a city $c$ as a list $w_c=1,\dots,n$, where $n$ is an integer corresponding to the total number of tiles. We then denote the volume of traffic for app $\alpha$ in city $c$ at tile $i\in w_c$ at time $t$ on day $d$ as $\rho_{\alpha, c, i}(t, d)$.

In order to form a temporal description of the traffic volume, we first perform a spatial aggregation of the traffic for each app over the city for each day,

\begin{equation}
    \rho_{\alpha, c}(t, d) = \sum_{i \in w_c} \rho_{\alpha, c, i}(t, d),
    \label{time series eq}
\end{equation}
recovering a time series representing the total traffic in the city at each 15 minutes time interval $t$ for a day in the dataset. 

We utilize these time series to calculate baseline statistics against which to measure abnormal app traffic. Since  the fire of the Notre-Dame cathedral occurred on Monday April 15th, we construct the baseline and standard deviation based on app traffic from the weekdays, Monday through Thursday, preceding the fire in the dataset: March 18th to April 11th 2019. We exclude Fridays and weekends as we expect they will have more heterogeneous application traffic patterns. We also confirmed that no national holidays coincided with our baseline period. We denote the mean traffic volume over this period as

\begin{equation}
    \mu(\alpha,c,t) = \langle \rho_{\alpha, c}(t, d) \rangle_d .
\end{equation}
The value of $\mu(\alpha,c,t)$ at each time $t$ corresponds to the average traffic volume for the app $\alpha$ during the 15 minute window $t$. We also compute the standard deviation for each tile and time and denote it $\sigma(\alpha, c, t)$. For convenience, from here on we will drop the indices for app and city, and simply use $\mu(t)$ and $\sigma(t)$ to represent the average and standard deviation for any combination of app and city over the baseline period. Similarly, we will use $\rho(t,d)$ to refer to a time series on a specific day. \\

We now define the \emph{distance from baseline} at time $t$ on day $d$ as the following modified z-score,
\begin{equation}
    D(t,d) = \frac{\rho(t,d) - (\mu(t) + 2\sigma(t))}{\mu(t) + 2\sigma(t)},
\end{equation}
indicating the value of traffic volume in relation to two standard deviations from the baseline mean of traffic. 

Finally, we define a spike in the traffic of an app as a period of at least 3 consecutive time points $[s_{start},s_{end}]$ where $D(t)>0$, in order to account for noise. For symmetry, we define the end of a spike as a period of three consecutive time points $[s_{end}+1,s_{end}+3]$ where $D(t)<0$.
 
\subsection{Temporal Clustering}
Traffic volume spikes of applications exhibit different characteristics in terms of duration, amplitude and position in time. In this section we define a set of features on the $D(t)$ time series of an application and use these features as inputs to a clustering algorithm to characterize apps with similar traffic spiking patterns. We break this analysis into two days, clustering immediate reactions to the fire as application behaviors on the day of the fire and then looking at longer term behaviors of applications the day after the fire. We introduce an overnight gap on the day after the fire from midnight to $7:00$, since baseline weeknights have small variation compared to weekdays, meaning smaller standard deviations and more noise using our anomaly detection method. Considering that the smoke during the fire of Notre-Dame is first visible 18:43\footnotemark[1], we denote a spike occurring from $18:45$ of April 15th to the end of that day as spike 1, and a spike occurring from $7:00$ to $24:00$ of April 16th as spike 2. We compute at most one spike per day, with only a first possible spike being considered.  
The 5 features that we compute are:
\begin{itemize}
    \item ($s_i^{start}$, $s_i^{end}$): the starting and ending times of the $i$th spike
    \item $s_i^{duration}$: the duration of the spike
    \item $s_i^{max}$: the maximum value of $D(t)$ during the spike
    \item $s_i^{aggregate}$: The sum of all $D(t)$ values during the spike
\end{itemize}
We thus represent the traffic time series of each app on the day of and day after the fire as two features vectors of length 5, and we normalize the features vectors by the maximum and minimum values for spiking apps within each feature for the day. Finally, we perform a K-means clustering of the feature vectors. Given the difference between the number of apps that spike in Paris compared with the other cities and our interest in understanding if applications will behave different in the city of incidence of catastrophe, we cluster apps in Paris separately from all other cities. Thus, we compute clustering for the following groupings independently: Paris on April 15th, Paris on April 16th, other cities on April 15th, and other cities on April 16th. We choose number of clusters $k=5$ by computing the Sum of Squared Errors over values of $k$ and manually inspecting the curves. For a consideration of traffic spikes as a single event across both days as a 10 dimension features vector, see Supplementary Materials.

\subsection{Spatial Analysis}

\subsubsection{Traffic Spikes in Space}

In this section we investigate the spatial patterns of the increase in anomalous app traffic in response to the fire. Our spatial analysis is restricted to the traffic response to the fire of the application Twitter in the city of Paris. Paris is considered as this is the city of incidence of the fire, allowing for measurement with respect to the epicenter. Only Twitter is considered as this application experiences a both strong and long term response, giving the possibility of analyzing its behavior over time.

We first recall that the Netmob2023 Data Challenge dataset has a spatial resolution of 100$m^2$ given as a tiling of the city of interest. Hence, all spatial analysis can be conducted at this spatial resolution. Recalling the definition, in Equation \ref{time series eq}, of the volume of traffic, we will consider only the time series of $\rho_{\alpha, c, i}(t, d)$ for the city of Paris ($c=$Paris), and for Twitter ($\alpha=$Twitter). Now for \textit{each tile} $i\in w_{Paris}$ we define the relative Twitter time-series as $\rho_{i}(t, d)$.

We use these tile-wise time series to calculate a baseline mean and standard deviation for the traffic for each spatial tile. Analogously to what is presented in the previous sections, we construct our baseline statistics based on Twitter traffic on the weekdays Monday through Thursday from March 18th to April 11th, again excluding Fridays and weekends as we expect they will have different patterns. We denote the mean traffic volume over this period for a tile $i$ similarly as,

\begin{equation}
    \mu_{i}(t) = \langle \rho_{i}(t, d) \rangle_d .
\end{equation}
The value of $\mu_{i}(t)$ at each time $t$ now corresponds to the average traffic volume for Twitter during the 15 minute window $t$ in tile $i$ of the spatial grid of the city of Paris. We also compute the standard deviation for each tile and time and denote it $\sigma_{i}(t)$.

We can now define the tile-wise \emph{distance from baseline} at time $t$ on day $d$, in tile $i$ as the following modified z-score:
\begin{equation}
    D_i(t,d) = \frac{\rho_i(t,d) - (\mu_i(t) + 2\sigma_i(t))}{\mu(t) + 2\sigma(t)}.
\end{equation}
We then compute $D_i(t)$ for the app Twitter on each tile $i$ of the city of Paris starting from $t=$18:45 on April 15th, the day of the fire, until $t=$24:00. 

\subsubsection{Quantifying Spatial Effects}

We are interested in how information propagates across applications with respect to the source of the catastrophic event. That is: what spatial patterns emerge in response to catastrophic events, how can we quantify these patterns, and how do they evolve temporally? Since the concept of physical distance does not exist in the virtual application space, it is of particular interest to understand how quickly information propagates across the internet and how this translates to analogous physical spatial patterns.

In order to analyze physical spatial patterns we take inspiration from Collective Response of Human Populations to Large-Scale Emergencies by Bagrow et. al \cite{Bagrow_emergencies}, which considers anomalous traffic volume as it pertains to the distance from the epicenter of the catastrophe. We begin by constructing concentric square shells of increasing radial size centered on the epicenter, with radius being the diagonal of each square and the distance from the epicenter measured in km's. A shell is defined as the collection of all of the tiles included in the square of the current radius and excluding all of the tiles from the previous radial distance, in essence a square annulus. For each shell, we sum the total anomalous traffic, $\sum_{i \in r} D_i(t)$, within the shell and normalize by the number of tiles within the specified shell resulting in a measure that shows the mean tile-wise volume of application traffic as a function of the radius \cite{Bagrow_emergencies},
\begin{equation}
    \Delta D(r) = \cfrac{\sum_{i \in r_m} D_i(t)}{\sum_{r_m} i}, \forall m = 0,1,2,... M .
    \label{eq:aggregate measure}
\end{equation}

Additionally, we define radius zero to be the single tile at the epicenter of the event. In our analysis, we consider a maximum shell radius to be the value of the last tile before exiting the defined city tiling in any direction. This consideration is sufficient in our analysis to recognize spatial patterns, however, this directional consideration may be adjusted, particularly for events which may occur on the edge of a defined city tiling where radial shells will quickly exit the city in a certain direction.

At each radius, we now have a measure of abnormal spatial traffic which allows us to quantify changes in traffic at all radial distances. We posit that given Equation \ref{eq:aggregate measure}, radial outward spread from the epicenter would be represented with a decaying function, with the rate of decay pertaining to the intensity of the decay.

The authors in \cite{Bagrow_emergencies} generally consider change in anomalous traffic volume from the epicenter aggregated over some time period, giving the course-grained, general behavior of the response to catastrophe. However, since we are interested in how the radial spread of information relaxes, we consider the `instantaneous' function of Eq. \ref{eq:aggregate measure} at each measured time point. This more fine-grained approach to understanding radial spread shows how radial patterns evolve over time and relax from perturbation. Additionally, looking a individual snapshots of anomalous traffic volume ensures that the response of one measured time does not dominate an interval of measurement.

Now that we have a measure of radial spread that can be analyzed at each time point, we can examine how radial the spread of information is and how quickly the abnormal application traffic patterns relax. To do so, we begin by using linear regression to recover a best-fit slope for the data at each time-point, tracking how the slopes of the recovered fits change throughout the course of the event. The radial decay is stronger as the value of the slope decreases.

We further explore the comparison of the data spread and its changes to a radial spread using Kullback-Leibler (KL) divergence. We first measure the KL divergence of each instantaneous snapshot in the form of Eq. \ref{eq:aggregate measure} to the aggregated function of Eq. \ref{eq:aggregate measure} two hours after the event, showing how the data changes from before, during, and after the fire as compared to its two hour change immediately following the catastrophe.

We then extend our analysis by constructing a null model of optimal possible radial spread given the data. The null model is constructed for a time point by considering all values of abnormal traffic for spatial tiles in the maximum radius searched in that time point. These values are then sorted by maximum value outward from the epicenter. That is, the maximum value of all tiles in the snapshot is assigned to the epicenter of the event, the next $n$ highest values are then assigned to the $n$ tiles within the first radius considered away from the epicenter, and so forth. We can then calculate a function of the form of Eq. \ref{eq:aggregate measure} on this optimal radial spread according to the data. As a next step, we compare the true data distribution to the optimally radial null model at each point in time by measuring the KL divergence between the two distributions. Finally, we plot the measured KL divergences over time between the instantaneous snapshots to both the aggregate and null models, tracking how the spread of information approximates a radial spreading pattern and relaxes from the radial state over time.

\section{Results}\label{sec3}

\subsection{Anomalous Application Traffic and Behavioral Clusters}

The aim of this work is to capture how the usage, or \emph{volume of traffic}, of different applications varies in response to a catastrophic event and whether changes in traffic volume within the city of Paris are different from those found in other cities for the fire of Notre-Dame. We investigate what patterns, both temporally and spatially, of application traffic response  emerge. First, we identify apps whose time series of traffic volume present an above-baseline increase in response to the fire. We consider the daily baseline of traffic volume for each app, for each 15 minutes interval of a day of the week (Monday to Thursday, from 00:00 until 23:45), as the average value of traffic volume computed over the same 15-minutes interval on all weekdays. We then define a measure of \emph{distance from baseline} $D(t,d)$, at time $t$ on day $d$ (see Methods for formal definition). We thus consider a spike in the traffic of an app a time-span of at least $3$ consecutive time points where $D(t,d)>0$. 

In Table \ref{table:apps_list}, we list the 11 apps that present spikes of traffic in response to the fire in the city of Paris; we also identify subsets of the same apps whose traffic volume spikes above baseline in the cities of Lyon, Marseille, Montpellier, Rennes and Strasbourg. Abnormal app traffic volume spikes, both in Paris and in other French cities, are consistent with the assumption that human activity spreads across social-media as people react in real-time to catastrophic events. Interestingly, the applications spiking in cities other than Paris are always a subset of the applications that spike in the French capital, the city of incidence.

\begin{table}[h!]
\centering

\caption{Comprehensive table of all apps that spike in Paris (Pa), Marseille (Mrs), Lyon (Ly), Montpellier (Mtp), Rennes (Rn) and Strasbourg (Sg).}\label{tab:all_apps}
\begin{tabular}{lcccccc}
\toprule%
Application & Pa & Mrs & Ly & Mtp & Rn & Sg \\
\midrule
Apple Video & \checkmark &    & \checkmark  & \checkmark  &   & \checkmark  \\
 Apple iCloud & \checkmark &  &  &   & &  \\
 DailyMotion & \checkmark &  &  & \checkmark  & & \checkmark\\
 Facebook & \checkmark &  &  &  \checkmark & & \checkmark\\
 Facebook Live & \checkmark &  &  &   & & \\
 Facebook Messenger & \checkmark & \checkmark & \checkmark & \checkmark  &\checkmark &\checkmark  \\
 
 Instagram & \checkmark &  & \checkmark & \checkmark & &  \\
 
 Molotov & \checkmark &   &\checkmark& \checkmark& & \checkmark  \\
 
 Periscope & \checkmark & \checkmark & \checkmark & \checkmark  & \checkmark & \checkmark \\
 
 Twitter & \checkmark & \checkmark & \checkmark & \checkmark  & \checkmark & \checkmark \\
 
 WhatsApp & \checkmark &  &  & \checkmark  & &  \\
\botrule
\end{tabular}
\label{table:apps_list}
\end{table}

We aim to identify groups, or clusters, of similar application traffic spiking patterns that correspond to various responses to the event or shed light on the ways in which the news of the fire is spread across platforms. This includes both the short and long term response to the incident, i.e., in the hours immediately following the fire of Notre-Dame on April 15th and on the following day.
\subsubsection{Patterns on the Day of Fire}

\begin{figure}[H]
    \centering
    \makebox[\textwidth][c]{\includegraphics[width=1.4\textwidth]{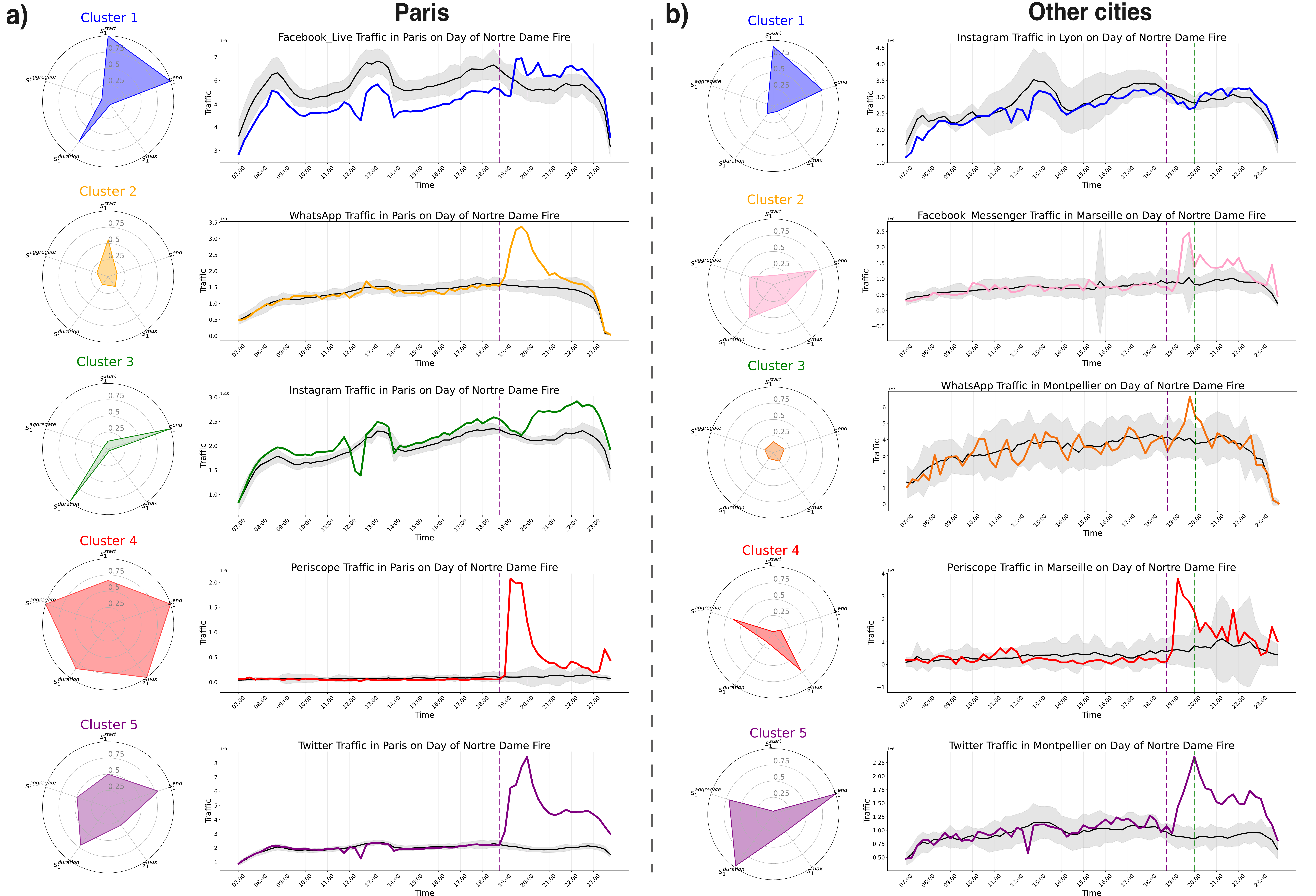}}
    \caption{\textbf{Clusters and Representative Time Series on Day of Fire:} \textbf{a)} For the city of Paris, the radar plots of clusters of applications are given on the left, showing the value of different features assigned to each cluster. On the right of each radar plot is a time series for an application that is representative of the applications in that cluster. Note, the information in the radar plots for each cluster is normalized within the city of Paris. The applications in each cluster are given in Table \ref{table:apps_list_cluster_paris_OF}. \textbf{b)} For all other cities being considered, the radar plots of clusters of applications are given on the left, with a representative time series of an of an application in the cluster shown on the right. Note, the information in the radar plots for each cluster is normalized for all cities excluding Paris. The applications in each cluster are given in Table \ref{table:apps_list_clusters_OTHER_of}.}
    \label{fig:day_of_together}
\end{figure}

First, we define five features that describe the spiking patterns observed in the individual traffic volume time series:

\begin{itemize}
    \item the starting time of the spike, $s_1^{start}$;
    \item the end of the spike, $s_1^{end}$;
    \item the maximum value of the spike, $s_1^{max}$;
    \item the length of the spike, $s_1^{duration}$;
    \item the sum of all values $D(t,d)$ in $[s_1^{start},s_1^{end}]$.
\end{itemize}
We then perform a K-means clustering  (see Methods for further details) on the spiking-features vectors of all apps that spike within the city of Paris (Figure \ref{fig:day_of_together}.a), and all the apps that spike in the other selected French cities (Figure \ref{fig:day_of_together}.b). For a full description of the values of the selected features for each application, see Supplementary Materials.

In Figure \ref{fig:day_of_together} each cluster consists of applications whose traffic time series display similar responses to the event as characterized by their features vector. We represent the typical response pattern of apps assigned to each cluster as a normalised radar plot of the \emph{centroid} of each cluster (Figure \ref{fig:day_of_together}.a, left). The radial axes of the radar plot correspond to the previously defined spiking features. The suffix $1$ in $s_1$ refers to \emph{first} day, or day of fire. We note that the spiking features are normalised with respect to its minimum and maximum values, across all apps, found for the city of Paris (Figure \ref{fig:day_of_together}.a), and for the other cities (Figure \ref{fig:day_of_together}.b).
Next to each radar plot, we show a representative time series of $D(t,d)$ of one of the applications assigned to the relative cluster (Figure \ref{fig:day_of_together}.a, right).

In the city of Paris, we recover from our clustering methods three clusters of applications with similar behavioral patterns and two outliers containing the singular behavior of an application. 
Cluster 1 is characterized by low-amplitude, long spikes with late start times relative to other clusters (high $s_1^{start}$), yet sustained until the end of the day (high $s_1^{duration}$ and $s_1^{end}$). Apps in this cluster are thus characterised by a delayed and moderate, yet sustained response to the spreading of the news of the fire. Cluster 2 is composed of apps with a late spike of relatively short duration and moderate amplitude. The response to the fire is thus closer to the collapse of the spire, rather than the moment in which the fire becomes visible, yet the perturbation of the traffic time series of apps in this cluster is the shortest w.r.t. the other clusters. Applications in Cluster 3 are characterized by long, low-amplitude spikes, with very early start times (low $s_1^{start}$). Apps in this clusters, such as Instagram, see a rise above baseline of their traffic volume as soon as the fire becomes visible from the outside of the cathedral; the traffic volume stays persistently above baseline throughout the rest of the day, yet the intensity of such anomalous activity is moderate.

The first outlier cluster, Cluster 4 includes one app with a distinctive behavior, Periscope. Periscope is a live video streaming app: its traffic time series has a massive spike (10 times larger than average over the previous weeks in the same time points) at 19:00, i.e., in the 15 minutes immediately following the smoke. The amplitude of the spike then decreases rapidly after 45 minutes, yet it stays above-baseline until the end of the day. This pattern of usage regarding Periscope is unique to the application, showing a strong preference in user interface.

Finally, the second outlier cluster, Cluster 5, contains only the discussion-based social media platform Twitter. The large, well sustained spike that characterizes Twitter traffic in Paris is particular to the response to catastrophe on this platform.

In other cities, we also recover 5 clusters of applications, giving further insight into user preference within and across cities. The applications and their respective clusters are illustrated in Table \ref{table:apps_list_clusters_OTHER_of}. Notice that although the same application in different cities mostly belong to the same clusters, this is not always the case. We also obtain one outlier cluster which contains a single application, Instagram in Lyon, which demonstrates completely distinct behaviour from both other application usage in Lyon, and Instagram usage in other cities. This emphasizes the conclusion that variations in application traffic behavior is strongly related to city-specific user preferences, and usage dynamics.

The app in Cluster 1, Instagram in Lyon shows a similar, delayed spike as Cluster 1 in Paris, yet the duration of such above-baseline traffic is very short. Cluster 2 comprises apps characterised by early and sustained spikes, that however relax back to baseline well before the end of the day.
Cluster 3 groups together apps whose traffic is perturbed early in response to the fire, whoever the amplitude and duration of such spikes are very low in comparison with apps in the other clusters.
Cluster 4, which includes the application Periscope for almost all other cities, experiences a short but intense spike of large amplitude, coherently with what is observed in Paris. Finally, Cluster 5 is composed of applications experiencing elevated levels of application traffic that last until the end of the day.

While these patterns have slight variations between the 5 types of behavior in Paris and the 5 types of behavior in other cities, there are notable similarities in user usage for applications in response to catastrophe. This is particularly relevant to the outlier clusters in the city of Paris and how they translate to the across-city clusters. Both in Paris and in other cities, the live video streaming app Periscope is characterized by a very large, yet brief, spike in usage, taking place shortly after the start of the fire. Like in Paris, a single cluster (Cluster 4 of other cities) is characterized only by application traffic patterns of the app Periscope across 4 other cities. This result shows the consistency of user interaction with this specific platform, showing that user tendencies can conform across for an application with a narrow scope, such as live video streaming. Patterns that are similar the French capital and in the other cities, are also evident in the behavior of Twitter. 4 of 5 twitters cluster into Cluster 5, however participation in this cluster is shared with other apps, showing general trends as well as user preference effects. Such application experiences a large spike after the start of the fire, has a sharp increase in traffic until the collapse of the spire of the cathedral, and remains persistently well-above baseline during the rest of the day for Paris as well as for all other cities. For both Paris and other cities there are also two clusters (Cluster 1 in both Figures \ref{fig:day_of_together}.a and \ref{fig:day_of_together}.b, Cluster 2 in Paris and Cluster 3 in other cities)  with shallow spikes that differ by their starting times, that is soon after the fire or relatively late with respect to other applications. 

We further consider whether applications in the other cities, excluding Paris, cluster similarly across these cities. We find that while this is mostly the case, this phenomenon is not exclusive. Applications that spike in multiple cities tend to cluster together, for example, Facebook Messenger clusters together for all the cities. However, there is some slight variation of application usage based on within-city user behavior. This is evident in that there are apps that spike in some cities, and not others. For example, Instagram and Molotov spike in some cities, but in all. Furthermore, there is some slight variation of clustering, such as how Periscope primarily clusters into Cluster 4, yet Periscope in Rennes is in Cluster 3. The same can be seen with Daily Motion, which appears for only 2 of the other 3 cities considered, in separate clusters. These results suggest the strong affect of user-preference.

We find that applications with the same \emph{`function'}, such a social media platforms, can exhibit very different response patterns. One might have better success in predicting app response patterns by a more fine grained type such as, messaging, broadcasting, image sharing, live video, etc. For example, the reaction-based platforms Facebook Live and DailyMotion cluster together. 

\begin{table}[h!]
\centering
\caption{Comprehensive table of all apps per cluster, in Paris on the day of fire, 15th April 2019. See Figure \ref{fig:day_of_together}. }\label{tab:paris_fire_apps}
\begin{tabular}{ccccc}
\toprule%
Cluster 1 & Cluster 2 & Cluster 3 & Cluster 4 & Cluster 5\\
\midrule
DailyMotion & Apple Video & Apple iCloud & Periscope  & Twitter \\
Facebook Live & Molotov & Facebook &  &\\
  & WhatsApp &Facebook Messenger&  &   \\
 & & Instagram & \\ 
\botrule
\end{tabular}
\label{table:apps_list_cluster_paris_OF}
\end{table}

\begin{table}[h!]
\centering
\caption{Comprehensive table of all apps per cluster spiking on the day of fire, 15th April 2019, for all other cities: Marseille (Mrs), Lyon (Ly), Montpellier (Mtp), Rennes (Rn) and Strasbourg (Sg). See Figure \ref{fig:day_of_together}.}\label{other_fire_apps}
\begin{tabular}{ccccc}
\toprule%
Cluster 1 & Cluster 2 & Cluster 3 & Cluster 4 & Cluster 5\\
\midrule

Instagram Ly & Facebook Messenger Mrs &Apple Video Sg & Periscope Mrs  & DailyMotion Mtp \\

  & Facebook Messenger Sg &Apple Video Mtp& Periscope Sg &Facebook Mtp\\
  
  & Facebook Messenger Ly &DailyMotion Sg& Periscope Ly  & Instagram Mtp  \\
  
  & Facebook Messenger Mtp& Facebook Sg & Periscope Mtp& Twitter Mrs \\
  
  & Facebook Messenger Rn&Molotov Sg  & & Twitter Sg\\
  & &Molotov Ly & & Twitter Ly\\
  & &Molotov Mtp & & Twitter Mtp\\
  & &Periscope Rn & & \\
  & &Twitter Rn & & \\
  & &WhatsApp Mtp & & \\
  
\botrule
\end{tabular}
\label{table:apps_list_clusters_OTHER_of}
\end{table}

\subsubsection{Patterns on the Day after Fire}

\begin{figure}[H]
    \centering
    \makebox[\textwidth][c]{\includegraphics[width=1.4\textwidth]{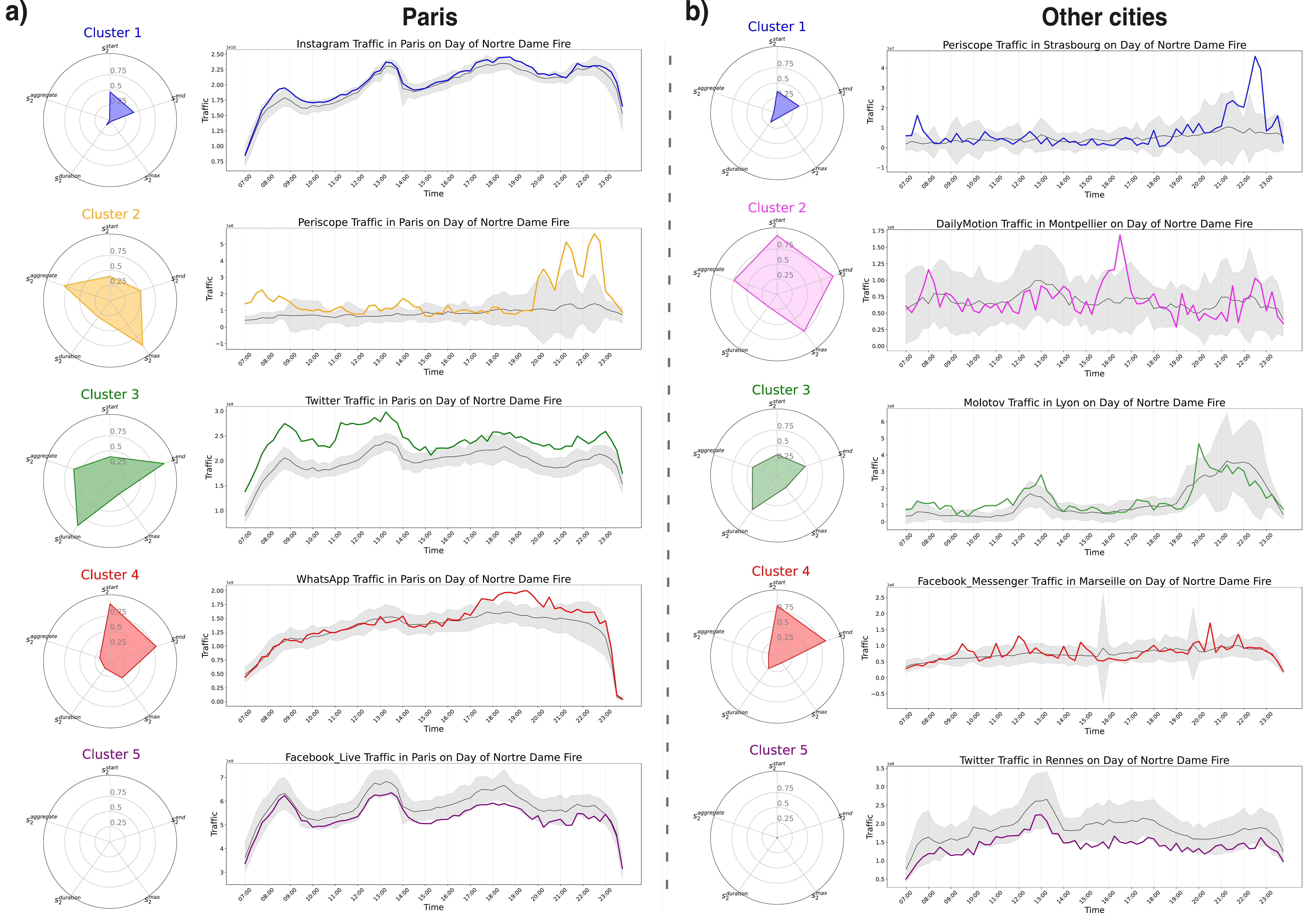}}
    \caption{\textbf{Clusters and Representative Time Series on Day after Fire:} \textbf{a)} For the city of Paris, the radar plots of clusters of applications are given on the left, showing the value of different features assigned to each cluster. On the right of each radar plot is a time series for an application that is representative of the applications in that cluster. Note, the information in the radar plots for each cluster is normalized within the city of Paris. The applications in each cluster are given in Table \ref{table:apps_list_cluster_paris_after}. \textbf{b)} For all other cities being considered, the radar plots of clusters of applications are given on the left, with a representative time series of an of an application in the cluster shown on the right. Note, the information in the radar plots for each cluster is normalized for all cities excluding Paris. The applications in each cluster are given in Table \ref{table:apps_list_clusters_OTHER_after}.}
    \label{fig:day_after_together}
\end{figure}

We now consider the long-term behavior of applications and their relaxation from abnormal traffic both in Paris and in other cities by considering app traffic on the day after the fire. We begin our analysis at 7:00 on the following day to account for small baseline standard deviations during the night. In Figure \ref{fig:day_after_together} we show app clustering for Paris (Figure \ref{fig:day_after_together}.a) and other cities (Figure \ref{fig:day_after_together}.b) on the day after the fire of Notre-Dame. We note that for most applications, there is a quick general relaxation of spiking behavior, that is, the abnormal app traffic does not last into the day after the fire. However, there are few exceptions, particularly in the city of Paris. The discussion based platform Twitter shown in Cluster 3 remains with elevated traffic levels throughout the second day, possibly due to new user content (discussion) being generated on the forum, unlike an image based platform like Instagram. Notably, there is some spiking behavior in clusters both for Paris and for other cities in the evening, as can be seen in Clusters 3, 4, and to a lesser extent, Cluster 1. This is particularly notable in the spiking of Periscope. We posit that this spiking, such as that of the video application Periscope and the messaging platform Whatsapp are associated to the organizing of and sharing of content concerning a vigil held in response to the fire at the Cathedral of Notre-Dame, which occurred approximately 24 hours after the fire and lasted into the night \cite{the_standard_vigil}. Additionally, we note that there no longer exist single-application clusters in Paris, as distinct behaviors collapse into more general long-term patterns. We conclude that for other cities, with the exception of Periscope in two cities, spiking patterns on the day after the fire are low and few, so as to be considered noise and too difficult to parse from general random spiking to conclude an association to the fire. Several applications do not exhibit long term behavior past the first night, as highlighted by Cluster 5 in both Paris and in other cities.

\begin{table}[h!]
\centering

\caption{Comprehensive table of all apps per cluster, in Paris on the day after the fire, 16th April 2019. See Figure \ref{fig:day_after_together}.}\label{paris_after_apps}
\begin{tabular}{lcccccc}
\toprule%
Cluster 1 & Cluster 2 & Cluster 3 & Cluster 4 & Cluster 5\\
\midrule

Apple iCloud & Molotov & Apple Video & DailyMotion  & Facebook Live \\

 Facebook& Periscope &Twitter& WhatsApp &Facebook Messenger\\
 
  Instagram&  &&  &   \\
  
\botrule
\end{tabular}
\label{table:apps_list_cluster_paris_after}
\end{table}

\begin{table}[h!]
\centering
\caption{Comprehensive table of all apps per cluster, on the day after the fire, 16th April 2019, in the other cities: Marseille (Mrs), Lyon (Ly), Montpellier (Mtp), Rennes (Rn) and Strasbourg (Sg).  See Figure \ref{fig:day_after_together}. }\label{other_after_apps}
\begin{tabular}{ccccc}
\toprule%
Cluster 1 & Cluster 2 & Cluster 3 & Cluster 4 & Cluster 5\\
\midrule

Periscope  Sg & DailyMotion Sg & Molotov Ly & Apple Video Ly  &  Apple Video Sg\\

 Periscope Rn & DailyMotion Mtp &Molotov Mtp& Apple Video Mtp &Facebook Sg\\
  
  & Facebook Mtp &Twitter Mtp& Facebook Messenger Mrs  & Facebook Messenger Sg\\
  
  & Periscope Mrs&  & Periscope Ly& Facebook Messenger Ly\\
  
  & Periscope Mtp& & & Facebook Messenger Mtp\\
  
  &  && & Facebook Messenger Rn\\
  & & & &Instagram Ly\\
  & & & & Instagram Mtp\\
  & & & &Molotov Sg\\
  & & & &Twitter Mrs\\
  & & & &Twitter Sg\\
  & & & &Twitter Ly\\
  & & & &Twitter Rn\\
  & & & &WhatsApp Mtp\\
  
\botrule
\end{tabular}
\label{table:apps_list_clusters_OTHER_after}
\end{table}

\newpage
\subsection{Spatial Response}

\begin{figure}[H]
    \centering
    \includegraphics[width=0.8\textwidth]{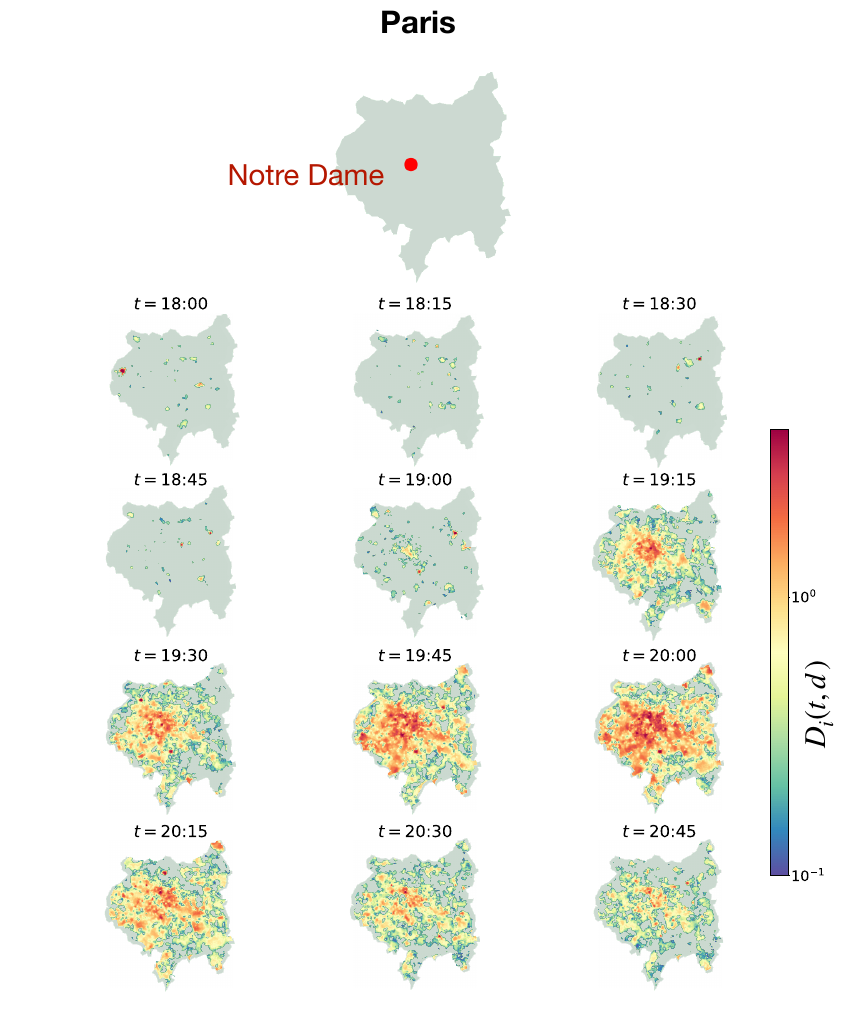}
    \caption{\textbf{Map of Paris with Abnormal Twitter Traffic.} The outline of the metropolitan area of Paris is shown at the top, with a red dot indicating the position of the Notre-Dame Cathedral. Below, each 15-minute timestamp shows the same city outline, now with each spatial tile as defined by the dataset displaying its corresponding tile-wise distance, $D_i(t)$, from baseline traffic. Only $D_i(t) > 0$ is considered. Recall that the first time-point considered to be related to the fire of Notre-Dame is 18:45.}
    \label{fig:Paris_spatial_spikes}
\end{figure}

The aim of this section is to investigate how information propagates across applications with respect to the geographic location of the source of the catastrophic event. We consider the association between information spread, in both time and intensity, and the distance from the epicenter of the catastrophe. We also aim to identify and quantify spatial patterns of above baseline traffic volume and their evolution over time. We define spatial above-baseline traffic, $D_i(t)$, tile-wise as the difference in application traffic volume in that tile from two standard deviations above its weekday mean (see Methods for formal definition).
For this analysis, we only take into account the application Twitter in the city of Paris, as it undergoes a sustained, abnormally elevated app traffic volume in the city of incidence. In Figure \ref{fig:Paris_spatial_spikes} we plot abnormally elevated application traffic volume, $D_i(t)$, in the metropolitan area of Paris on the day of the fire from 18:00 to several time points following the fire. We note how, before the fire, the above-baseline traffic is spatially distributed as random noise. However, after the fire starts, we detect an approximately  \emph{radial}, outward spreading pattern emanating outward from the Notre-Dame Cathedral. Such pattern appears to persist until end of the day, relaxing over time in both intensity and \emph{traveled} distance.

\begin{figure}[H]
    \centering
    \includegraphics[width=\textwidth]{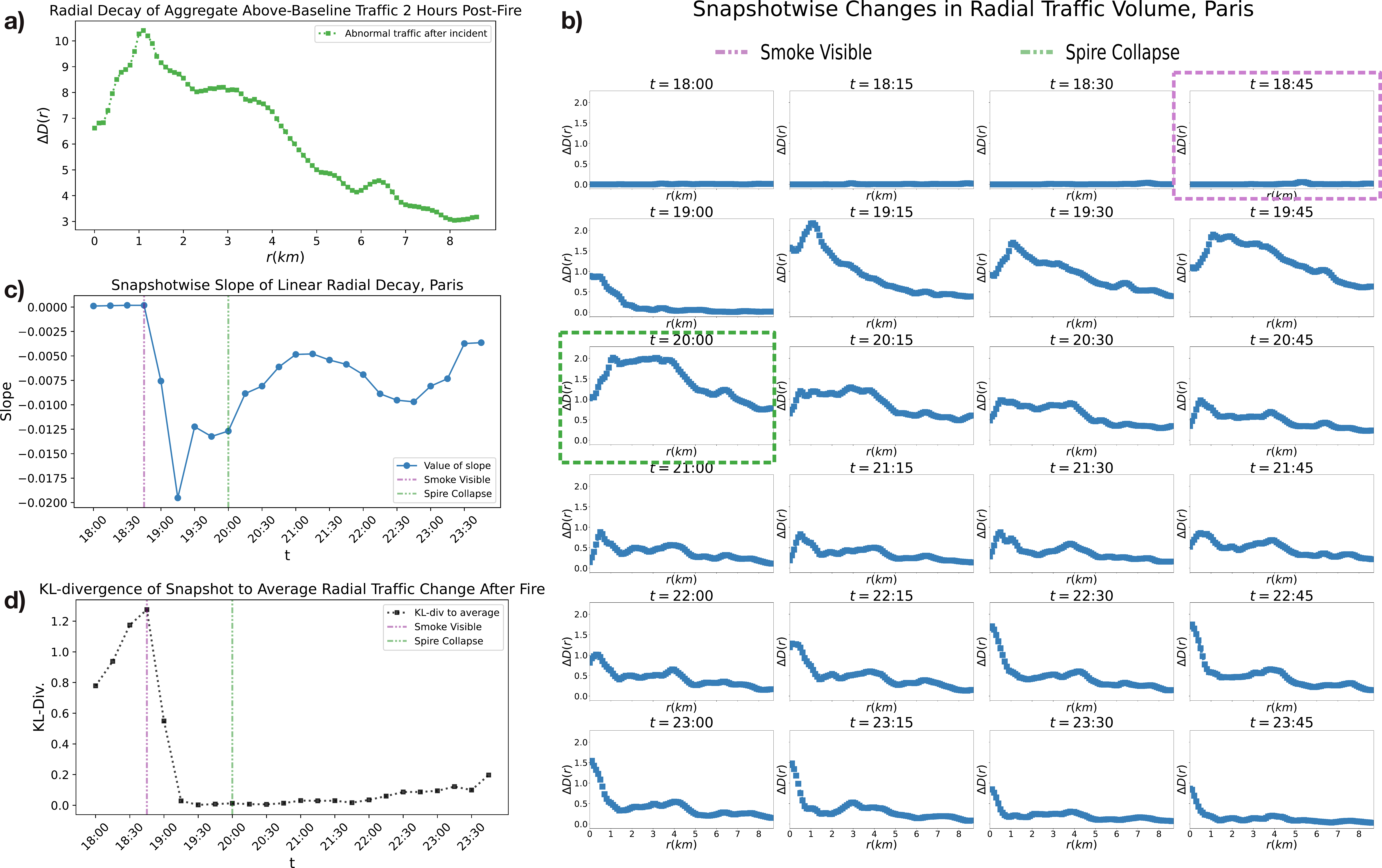}
    \caption{\textbf{Instantaneous vs Aggregate Radial Patterns in Paris in Response to the Fire in Paris:} \textbf{a)} The change in abnormal traffic volume as a function of radius from the epicenter of the catastrophe (see Equation \ref{eq:aggregate measure}) is shown over the course of two hours (between 18 : 45 and 20 : 45), showing the course grained, general behavior of spatial Twitter traffic response in the aftermath of the fire. The generally decreasing behavior of the function (particularly from 1km) implies a radial spread of information. \textbf{b}) Changes in abnormal traffic volume with respect to distance from epicenter are shown at each 15-minutes time interval. From these snapshots, we can then track changes in the ``radial-ity" of information spread. c) The slope of the best-fits line to each snapshot over time. This shows how the strength of radial spread changes over time. That is, a more radial distribution of information outward from the epicenter of the fire corresponds to a more negative slope, or a faster decay of the change in aggregate traffic from the epicenter. d) The Kullback-Leibler divergence of each snapshot in panel b to the average of snapshots from 18 : 45 to 20 : 45, showing how the strength of the radial spread of information varies with respect to its average change over the defined interval at each time-point. This emphasizes the change in traffic spread before and after the fire and the maintaining of the radial pattern throughout the rest of the night.}
    \label{fig:Paris_aggregate_fig}
\end{figure}

Given that using mobile phones, users at any physical distance could in theory access information at the same time, a pattern of radial spread that appears dependent on physical distance is unexpected. We thus quantitatively address the pattern of outward radial spread, with the cathedral as its origin. Analogously to what was done in \cite{Bagrow_emergencies}, we track how the intensity of anomalous traffic, aggregated over the course of 2 hours after the start of the catastrophe, changes with the distance from the epicenter, i.e., the cathedral. We use the change in traffic volume in subsequent squares of increasing diagonal length normalized by the number of tiles within the considered distance, $\Delta D(r)$,  to describe this relationship. That is, we analyze how aggregated traffic, $D(r)$, within some defined area changes as the radial distance from the epicenter increases (see Methods for further details). Noting that a radial spread of intensity of abnormal traffic outward from the epicenter would be approximated by a decaying function of this form (see Methods, Eq. \ref{eq:aggregate measure}), we find that in general the spread of information after the fire seems to follow a dependence on the radial distance from the epicenter (see Figure \ref{fig:Paris_aggregate_fig} panel a).

We can then inspect how the radial spread, or change in application traffic as a function of the radial distance from the epicenter of the catastrophe, evolves over time (see Figure \ref{fig:Paris_aggregate_fig} panel b). We note that in the time points before the fire, the spread of abnormal traffic spikes does not experience a dependence on the distance outward from the epicenter of the catastrophe. In the time-points immediately following the fire, there is a rapid shift in the dependency of the volume of traffic based on radius from epicenter of catastrophe (see Figure \ref{fig:Paris_aggregate_fig} panel b). We note that the more intense the decaying behavior of the functions in the form of Eq. \ref{eq:aggregate measure}, the more the distribution of data approximates a radial behavior. Thus, to further quantify the radial pattern of the data and how this radial pattern relaxes, we use linear regression to calculate a slope for each snapshot function shown in Figure \ref{fig:Paris_aggregate_fig} panel b. We can then plot how the slope of these best-fits lines change in time, with more negative slopes corresponding to more intense radial spread, which relaxes over time (see Figure \ref{fig:Paris_aggregate_fig} panel c).

We can also measure the KL divergence between the distribution of data in each snapshot of the form of Eq. \ref{eq:aggregate measure} and the mean of the these snapshot distributions in the two hours following the fire (see Figure \ref{fig:Paris_aggregate_fig} panel d). This describes how each snapshot behaves with respect to the average over the two hour interval, highlighting the striking difference between the behavior of the distributions before and following the fire, and emphasizing the persistence of the data pattern in response to the fire throughout the night. 

\begin{figure}[H]
    \centering
    \includegraphics[width=\textwidth]{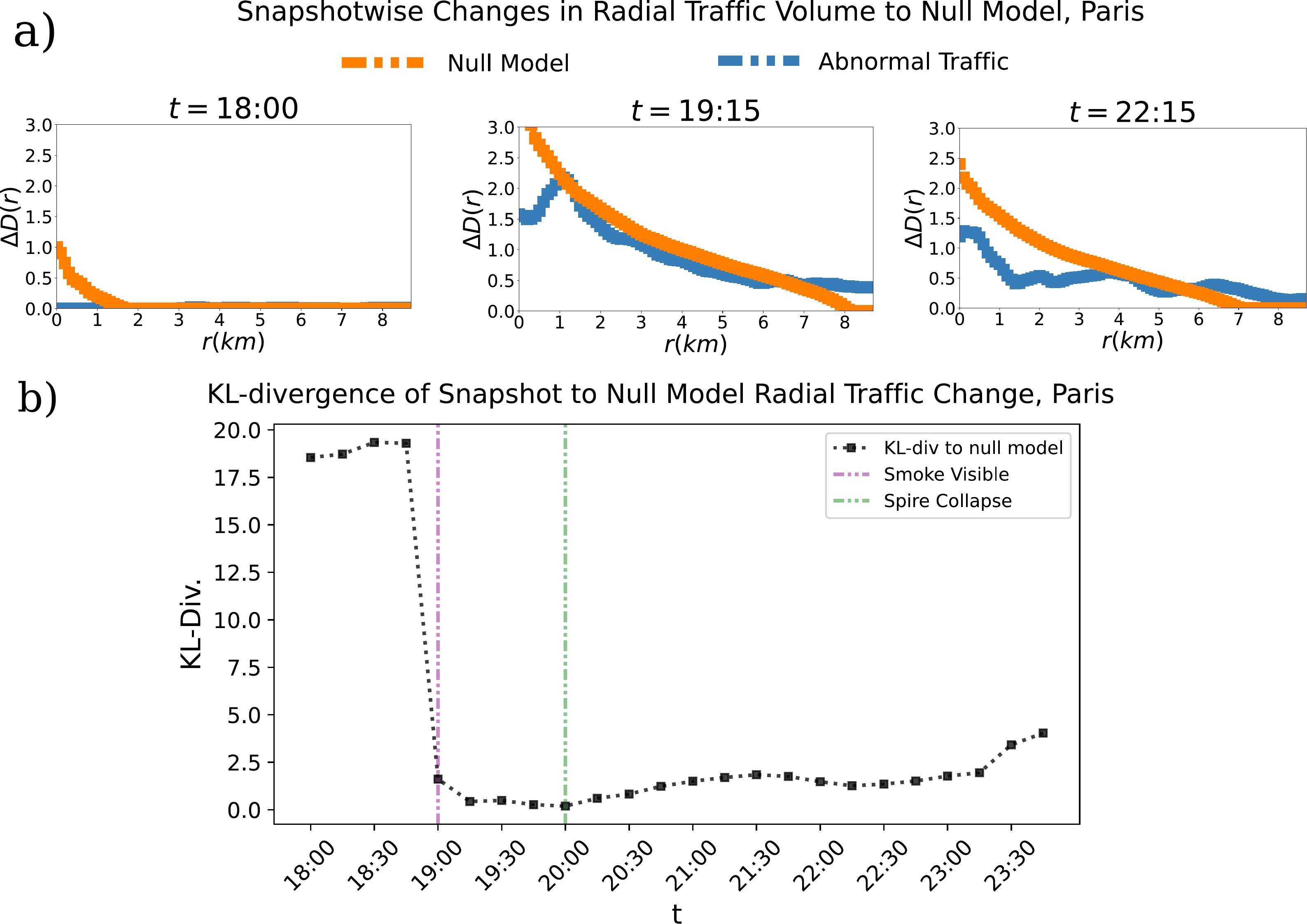}
    \caption{\textbf{Data vs. Null Model in Twitter Response to the Fire of Notre-Dame :} \textbf{a)} The change in abnormal traffic for the data and the null model at that respective time points is shown for 3 time points. Note that the first recorded time point after fire is 18 : 45. \textbf{b)} The Kullback-Leibler divergence of each snapshot to the null model of most radial possible spread constructed at that time point is shown.}
    \label{fig:Paris_null_fig}
\end{figure}

We then proceed to construct a null model according to the data of abnormal traffic distribution with optimally radial spread at each snapshot (see Methods). We can the measure the KL-divergence at each time-point between the function in the form of Eq. \ref{eq:aggregate measure} constructed from the data and the same functional form given by the optimally radial null model. This tells us how radial the reaction pattern in response to the catastrophe is with respect to its most radially possible distribution. We see that in the time points immediately preceding the fire, the distribution of the data is far from the distribution of the radial null model, showing a lack of dependence of abnormal traffic data from the epicenter of the fire. We then find a dramatic decrease in the divergence from the radial null model in the time points immediately following the fire, followed by a gradual rise in divergence, however, the approximation to maximum radial spread remains prevalent until the end of the day (see Fig \ref{fig:Paris_null_fig}).

This is an exemplar result, as one might have previously assumed that as it is possible to receive information of catastrophe at the same time, application traffic spikes would be distributed uniformly with respect to the epicenter. However, the information and its intensity spread outwards from the epicenter, like a real-world fire. This result is, in fact consistent with results from \cite{Bagrow_emergencies}, who found a similar radial spread when looking at call traffic volume in response to catastrophic events such as earthquakes. The traffic intensity in those cases radiates outward in an radial way and the function in the form of Equation \ref{eq:aggregate measure} aggregated over a two hour period hours approximates an exponential decay.

\begin{figure}[H]
    \centering
    \includegraphics[width=0.8\textwidth]{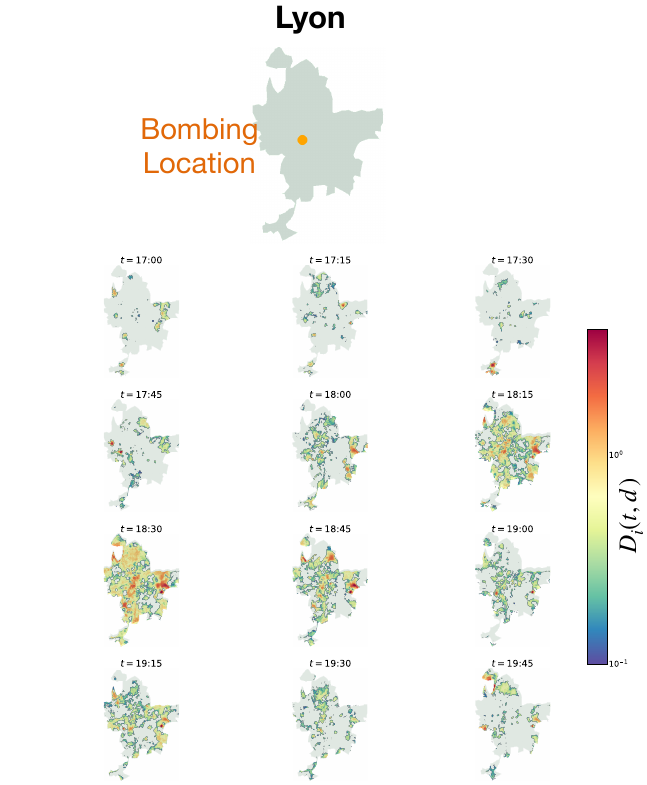}
    \caption{\textbf{Map of Lyon with Abnormal Twitter Traffic.} The outline of the metropolitan area of Lyon is shown at the top, with a red dot indicating the site of the bombing. Below, each 15-minute timestamp shows the same city outline, now with each spatial tile as defined by the dataset displaying its corresponding tile-wise distance, $D_i(t)$, from baseline traffic. Only $D_i(t) > 0$ is considered. Recall that the first time-point considered to be related to the bombing is 17:30.}
    \label{fig:Lyon_spatial_spikes}
\end{figure}
\subsubsection{Lyon, a Further Study of Spatial Spreading Patterns}
\begin{figure}[h!]
    \centering
    \includegraphics[width=\textwidth]{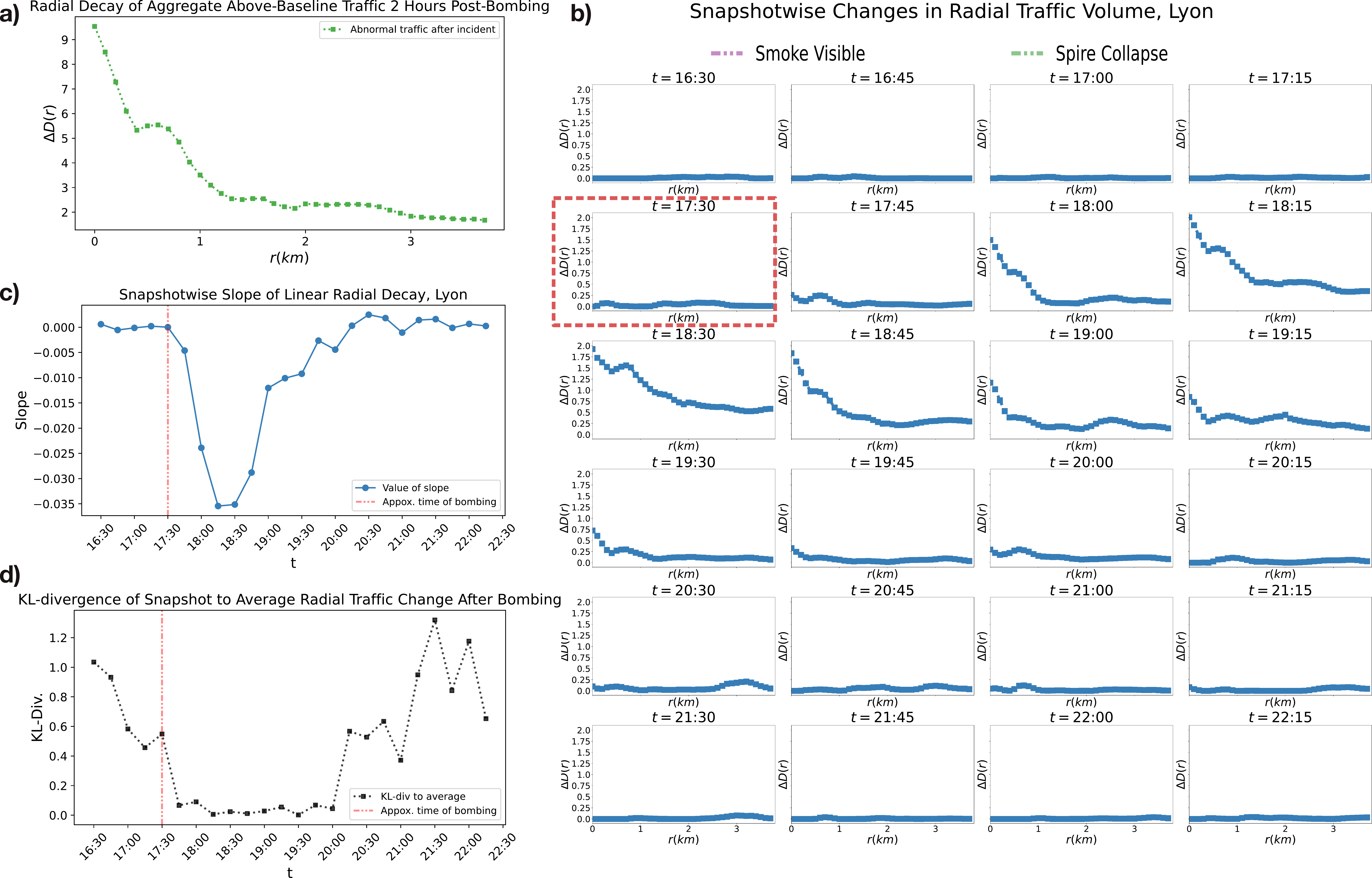}
    \caption{\textbf{Instantaneous vs Aggregate Radial Patterns in Twitter Response to Bombing in Lyon}\textbf{a)} The change in abnormal traffic volume as a function of radius from the epicenter of the catastrophe (see Equation \ref{eq:aggregate measure}) is shown over the course of two hours (between 17 : 30 and 19 : 30), showing the course grained behavior of spatial Twitter traffic response in the aftermath of the bombing. The generally decreasing function implies a radial spread of information. \textbf{b}) Changes in abnormal traffic volume with respect to distance from epicenter are shown at each 15-minutes time interval. c) The slope of the best-fits line to each snapshot over time. This shows how the strength of radial spread changes over time. d) The Kullback-Leibler divergence of each snapshot in panel b to the average of snapshots from 17 : 30 to 19 : 30, showing how the strength of the radial spread of information varies with respect to its average change at each time-point.}
    \label{fig:Lyon_aggregate_fig}
\end{figure}

This section focuses on a second catastrophic event, a bombing in Lyon, France, occurring within the time period included the NetMob2023 dataset. We extend our spatial analysis of spiking traffic volume of the application Twitter in response to this event in order to ascertain if there is consistency within the spatial spreading pattern results and to understand what differences may arise. 
The Lyon bombing occurs at approximately 17:28 on Friday, May 24th of 2019 \cite{abc_news_bombing}. Consistent with our analysis of the fire of Notre-Dame in Paris, we focus only on the abnormal traffic of the application Twitter. We plot abnormally high application traffic volume in Lyon over several time points in Figure \ref{fig:Lyon_spatial_spikes}. Abnormally high traffic volume is measured with respect to baseline statistics calculated from all Friday's prior to the bombing within the dataset.
Unlike in the Twitter traffic in Paris during the Notre-Dame fire, it is not plainly evident by inspection a clear radial spatial spread outward from the epicenter of the catastrophe. Therefore, we apply our previous spatial analysis methods to more thoroughly describe the spatial propagation of information (see Figure \ref{fig:Lyon_aggregate_fig}). 

Consistent with results from spatial methods applied to the fire of Notre-Dame, we find that in the two hours following the catastrophic event above-baseline traffic volume also experiences a dependence on the radial distance with respect to the epicenter of the catastrophe (see Figure \ref{fig:Lyon_aggregate_fig} panel a). 

 We look at this pattern at each time-point to further examine the reaction of application traffic to the event and assess the excitation and relaxation of information spread. Taking functions of the form of Eq.\ref{eq:aggregate measure} for each time point, we can see the appearance of a radial pattern after the event, followed by a quick relaxation of this radial pattern in time (see Figure \ref{fig:Lyon_aggregate_fig} panel b). We can quantify this as before with slopes of linear regressions on each snapshot, once again demonstrating the evolution of the intensity of radial spread in the time following a catastrophe and its relaxation in time (see Figure \ref{fig:Lyon_aggregate_fig} panel c). Finally, we calculate the KL divergence between the data distribution in each snapshot according to Eq.\ref{eq:aggregate measure} and the mean of these distributions in the two hours after the bombing (see Figure \ref{fig:Lyon_aggregate_fig} panel d). The KL divergence of the data to the 2 hour average is far noisier than the catastrophe of the fire of Notre-Dame. While there is consistency with the mean in the time-points immediately following the fire, the duration of this consistency is far shorter that of the catastrophe in Paris. We posit that the changes to the duration of the radial spread is due to the different duration of the two events. That is, the cathedral fire occurs visibly for several hours, including the collapse of the historic spire and with continuous updates as firefighters battled the blaze. On the other hand, the event of the Lyon bombing was singular and not as long lasting.

\begin{figure}[H]
    \centering
    \includegraphics[width=\textwidth]{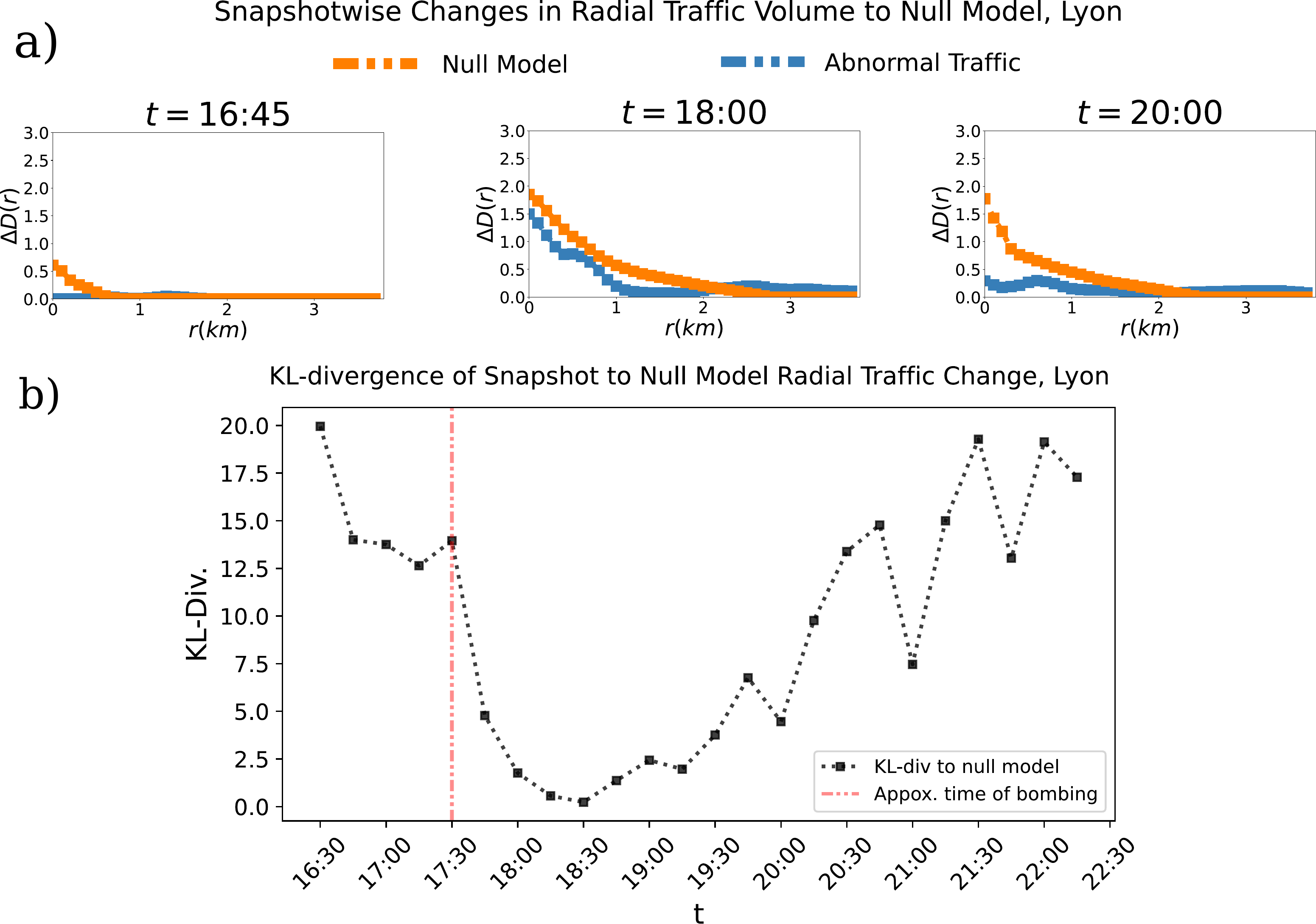}
    \caption{\textbf{Data vs. Null Model in Twitter Response to Bombing:} \textbf{a)} The change in abnormal traffic for the data and the null model at that respective time points is shown for 3 time points. Note that the first recorded time point after the bombing is 17 : 30. \textbf{b)} The Kullback-Leibler divergence of each snapshot to the null model of most radial possible spread constructed at that time point is shown.}
    \label{fig:Lyon_null_fig}
\end{figure}

We then construct a optimally radial null model according to previous methods and measure the KL divergence at each time point between the data and the null model to asses how closely the response pattern to the catastrophe represents a spatially radial spread (see Figure \ref{fig:Lyon_null_fig}). We find that following the bombing, the data pattern closely approximates the null model of radial distribution. These results are consistent in terms of radial spread of information after a catastrophe, between the two cities. However, the different duration over which the two events unfold drastically affect the time of relaxation to the generally noisy baseline of the traffic volume after the two events. 

\begin{figure}[H]
    \centering
    \includegraphics[width=1\textwidth]{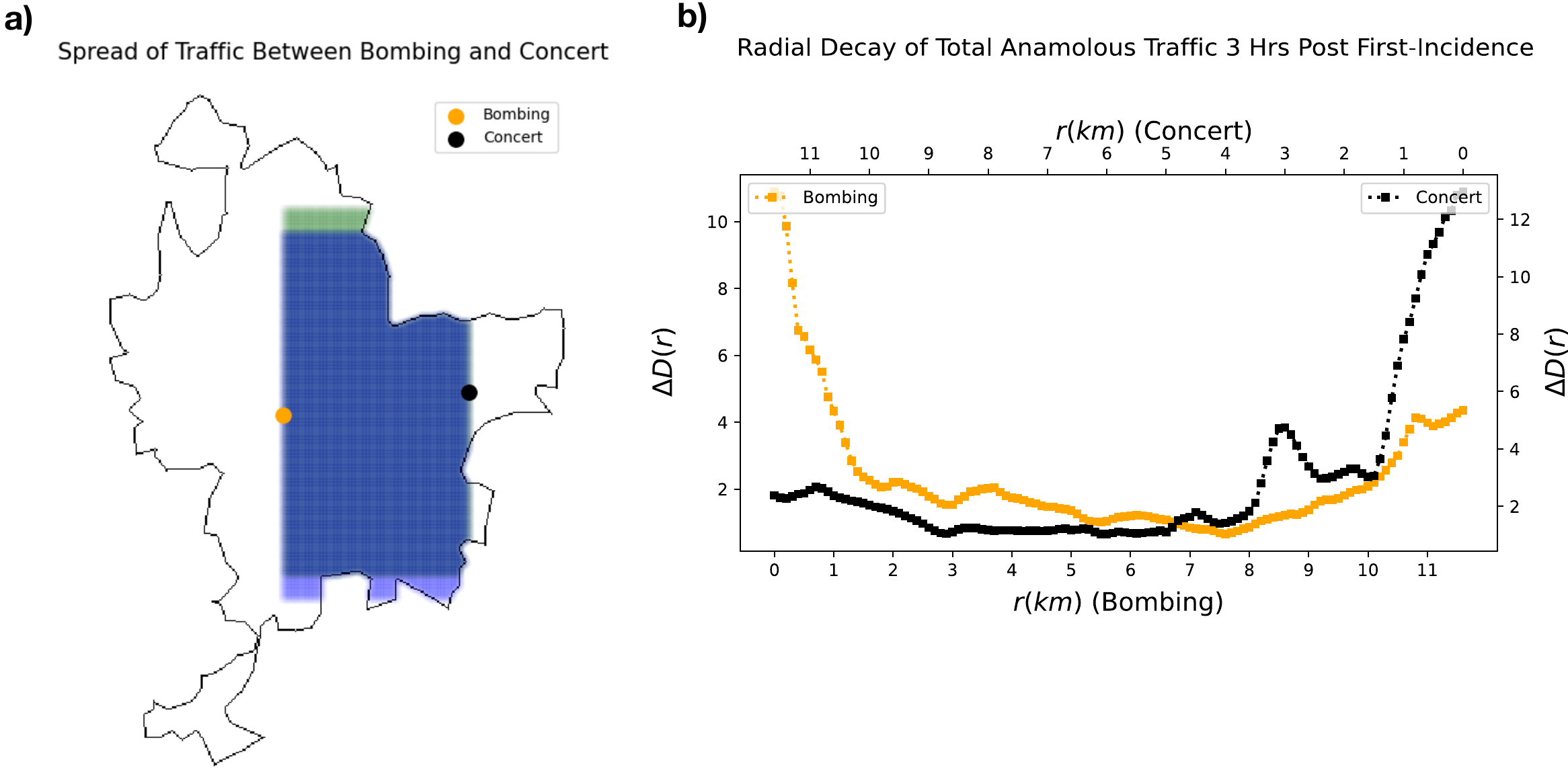}
    \caption{\textbf{Traffic overlap between bombing and concert in Lyon} \textbf{a)} The outline of the metropolitan area of Lyon is shown, with a orange and black dots indicating the locations of the bombing and concert, respectively. The abnormal traffic between these two location \emph{within} the city is considered, using squares of increasing radial size. \textbf{b)} The change in aggregate of abnormal traffic between the two epicenters of both the bombing and the concert are shown, from both epicenters towards the other epicenter.}
    \label{fig:Overlap_lyon}
\end{figure}

Finally, we note that on the day of the catastrophe, there was a large planned event in Lyon, an Ed Sheeran concert, which was advertised to begin at 18 : 00 \cite{billeto_concert}. In Figure \ref{fig:Overlap_lyon}, we demonstrate the extension of the methods of abnormal application traffic and pattern detection to planned events, and the interaction between two events. We note that we slightly modify our radial shell method to consider only the tiles \emph{between} the two events \emph{within} the city, until reaching the other event. Our results suggest a slight interaction between the traffic of the two events, however, this is mitigated by the large distance between the 
two events.

\section{Discussion}\label{sec12}

In this work we analyze time series of spatially distributed mobile phone applications and investigate the deviation from baseline activity in such traffic triggered by a catastrophic event, i.e., the fire of the Notre-Dame cathedral that took place on Monday, 15th April 2019. We use the Netmob2023 Data Challenge Dataset \cite{netmob23} to investigate patterns of use across applications, look into how these patterns might vary within the city of incidence as opposed to other cities, and examine how the information propagates in space and time. We are particularly interested in how temporal and spatial patterns of anomalous application traffic arise and relax over time.

We introduce a definition of \emph{spike} (see Methods) of anomalous application traffic and detect the presence of such spikes for apps in the cities of Paris, Marseille, Lyon, Montpellier, Rennes and Strasbourg, on the day of and the day after the fire of Notre-Dame. We note that applications that spike in other cities are a subset of applications that spike in Paris.

The categories of apps whose time series of distance-from-baseline traffic have at least one spike include: social media, video streaming, and messaging platform. Interestingly, applications with the same function, such as social media, may exhibit differences in behavior. For example, Instagram follows a similar behavior patter across all cities, characterized by short (low duration) traffic spikes across all cities. However, Twitter, another social media network, is characterised by long and sustained spikes in traffic in time (see Figure \ref{fig:Fig0}). This could be due to the tendency of users to rely on Twitter more than Instagram as a source of news and information on ongoing events \cite{lehmann_twitter_cluster}.

To further investigate these differences in the above-baseline traffic of apps of similar function, we investigate how applications cluster together based on the spiking patterns in time over the interval of interest. We separate the day of incidence and the day after to best capture the long term behavior of applications and how these may differ. We also perform the clustering separately for apps that spike in Paris and apps that spike in the other 5 cities to further investigate how patterns may differ in a city of incidence.

We find that between Paris and other cities on the day of interest, the applications experience generally similar trends in clustering, however, there is some diversity in the spiking patterns. For example, in application clustering can be heavily influenced by user-preference within a city. Further since broad application categories, such as social-media, cannot fully capture the diversity of application traffic response, it seems that we may need a more fine-grained look at function of applications, such as broadcasting and discussion based labeling. Distinct behaviors can be seen in Live Video streaming platforms, where short but intense spikes are seen. Also notable is the discussion based platform Twitter, which experiences a high and sustained spike. Most applications experience a fast relaxation from spiking, given that the next day many applications do not experience spikes in application traffic above baseline activity. This is especially true in cities that are not the city of incidence, showing that general interest in the event is more strongly maintained in the city of incidence. 

The clustering of the time series of distance-from-baseline of app traffic in the other cities highlights how the effect of the event on an app is differs across cities. Nonetheless, we note how live-streaming and, in general, video streaming apps such as Periscope, DailyMotion and Molotov, while with a delay with respect to Periscope in Paris, display a similar early spiking trend. We speculate that the delay in time of the anomalous traffic activity could be correlated with the physical distance from Paris, as well as to the time it takes for the video and pictures of the flaming cathedral to be noticed on these platforms in other cities.

The longest lasting, most sustained spikes tend to occur in Twitter in all of the cities. Such a result suggests that Twitter might be a suitable candidate to investigate the spatio-temporal spreading of the event-triggered anomalous traffic in the city of Paris. We thus investigate the existence of a relationship between how information propagates in the virtual information sphere, and how such propagation unfolds in physical space. We find that the anomalous, above-baseline activity emanates outward of in the immediate proximity of the cathedral of Notre-Dame. We further compare the data to a null model of optimal radial spread, demonstrating the shifts to radial data patterns in response to the fire and the relaxation from radial patterns over time. This result is surprising in that the intensity of the spread appears to decay radially with the distance from the epicenter of the catastrophe, which is not immediately evident given the means of communication being investigated, that, in principle, do not rely on physical distance. This result is consistent with other studies showing that call traffic volume radiates outward from the epicenter of catastrophe \cite{Bagrow_emergencies}. We further explore the spatial spread and its relaxation through a further analysis of a bombing in the city of Lyon. We find consistent results, however, the relaxation of abnormal traffic patterns in Lyon is much quicker than the fire of Notre-Dame in Paris, likely due to the much shorter duration of the event.

There are a number of possible reasons for the radial spread effect we observe here. For example, Twitter posts are often geo-tagged, and may suggest posts to users based on their location. Furthermore, the structure of the social network itself may have a geo-spatial relationship, as social networks have been shown to be influenced by both geographic proximity and residential segregation~\cite{xu2022beyond}. Increased data granularity and information about the social network structure would be necessary for a deeper understanding of this radial spread effect. Finally, we show planned events may also exhibit radial spreading patterns and suggest that events may interact with each other.

In conclusion, we provide a description of how catastrophic events, particularly the fire of the Notre-Dame cathedral perturb the temporal and spatial patterns of app usage traffic. An interesting development of the current research would be to characterise the spreading of the above-baseline activity at a coarser spatial scale, thus considering entire cities as spiking tiles. Then, a definition of spiking delay with respect to Paris could be introduced, in order to investigate whether any spatio-temporal patterns could be found. We speculate that such result could be achieved with a more granular time resolution, as, potentially, other means of communication, such as news broadcasting, could enforce a synchronisation of the spiking activity in the other French cities.

\bmhead{Funding Information}
This project has been supported by the EPSRC grants EP/V013068/1, EP/V03474X/1, EP/Y028872/1, EP/W523781/1, and EP/T517653/1. This work was supported by The
Alan Turing Institute’s Enrichment Scheme. RS is supported by the Mathematical Institute Scholarship at the University of Oxford.

\bmhead{Availability of data and materials}
Access to the NetMob23 dataset was granted under a licensing agreement as part of the NetMob 2023 Data Challenge. Further information is available at https://netmob2023challenge.networks.imdea.org/. A full description of the dataset used in this work can be found in \cite{netmob23}.

\bmhead{Authors Contribution}
SM contributed to the conception and design of the work; to the analysis and interpretation of data; to the creation of new software; to the drafting and revision of the manuscript. NP contributed to the conception and design of the work; to the analysis and interpretation of data; to the drafting and revision of the manuscript. SAB contributed to the conception and design of the work and to the revision of the manuscript. RS contributed to the creation of new software used in the work and to the revision of the manuscript. TL contributed to the drafting and the revision of the manuscript. RL contributed to the conception and design of the work and to the drafting and revision of the manuscript.

\bmhead{Competing Interests}
The authors declare to have no competing interests.

\bibliography{sn-bibliography}

\newpage
\section{Supplementary Materials}
\subsection{Applications under consideration}

Our analysis considers the applications represented in the NetMob2023 Data Challenge dataset. The description of the 68 available applications is given in \cite{netmob23}. We first sort for relevance, with our criteria for included apps consisting social media, messaging, transportation and location, video-sharing and streaming, and cloud storage apps.

We recover the following list of 24 applications that will be assessed for traffic spikes:

\begin{multicols}{2}
\begin{itemize}

    \item Apple App Store
    \item Apple Video
    \item Apple iCloud
    \item Apple iMessage
    \item DailyMotion
    \item Facebook
    \item Facebook Live
    \item Facebook Messenger
    \item Google Drive
    \item Google Maps
    \item Google Play Store
    \item Instagram
    \item Molotov
    \item Orange TV
    \item Periscope
    \item Snapchat
    \item Telegram
    \item Twitch
    \item Twitter
    \item Uber
    \item Waze
    \item Web Transportation
    \item WhatsApp
    \item YouTube

\end{itemize}
\end{multicols}

Of the applications considered, 11 of 24 applications experience spike. The full table of all spiking app is found in Table \ref{table:apps_list} in the main text. None of the transportation or location applications considered experience abnormally high traffic spikes. Other applications that do not experience traffic spikes include the messaging app Telegram and the video streaming app Youtube. 

\subsection{Feature Values}

We include features that describe the commencement time, duration, and intensity of the application traffic spike. This includes the start and end time of the each anomalous traffic spike, as well as the magnitude of deviation from baseline traffic (see Methods for more details).

\subsubsection{Features, day of fire}
\begin{figure}[h!]
    \centering
    \includegraphics[width=\textwidth]{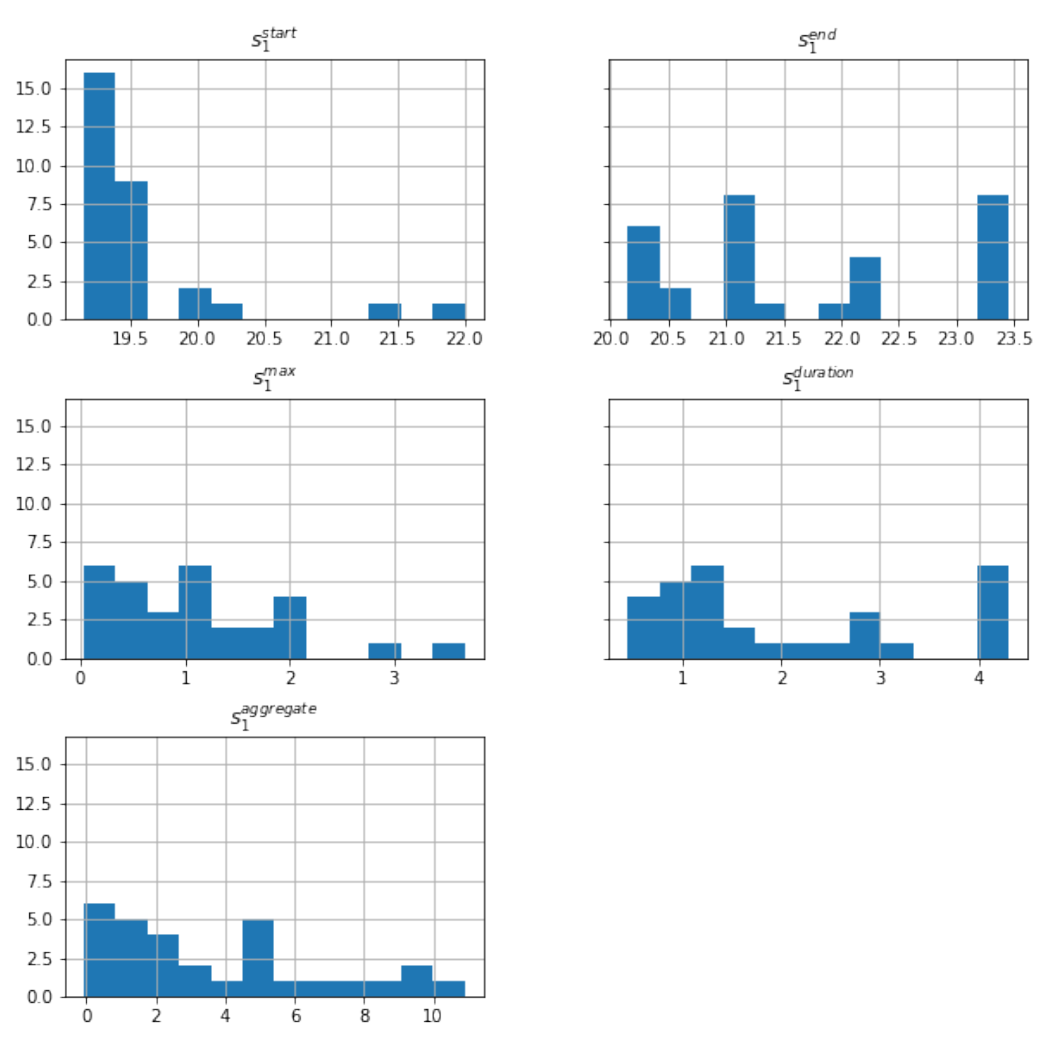}
    \caption{\textbf{Distributions of feature values}, The distribution of the values of each feature for all other cities on the day of the fire is shown.}
    \label{fig:Dist_feat_other_day_of}
\end{figure}

A full description of the non-normalized values for the features of spiking applications on the day of the fire of Notre Dame is given in Tables \ref{table:features Paris day of} and \ref{table:features other cities day of}. Table \ref{table:features Paris day of} shows the non-normalized values for features of applications in the city of Paris and Table \ref{table:features other cities day of} shows the values for the features of spiking applications in other cities. The distribution of the values of each feature across all applications is other cities on the day of the fire is shown in Figure \ref{fig:Dist_feat_other_day_of}.

\begin{table}[h!]
\centering
\begin{tabular}{|c c c c c c|} 
 \hline
                     App &  $s_1^{start}$ &  $s_1^{end}$ &  $s_1^{max}$ &  $s_1^{duration}[hrs.mins]$ &  $s_1^{aggregate}$ \\
 \hline
       Apple Video &     19.15 &   21.15 &    2.10 &         2.00 &          8.42 \\
      Apple iCloud &     18.45 &   23.45 &    0.40 &         5.00 &          3.54 \\
        DailyMotion &     19.30 &   23.45 &    1.02 &         4.15 &          8.05 \\
          Facebook &     19.00 &   23.45 &    0.32 &         4.45 &          3.52 \\
     Facebook Live &     19.30 &   23.45 &    0.11 &         4.15 &          0.51 \\
Facebook Messenger &     19.15 &   22.45 &    3.54 &         3.30 &         15.29 \\
         Instagram &     18.45 &   23.45 &    0.24 &         5.00 &          2.71 \\
         Molotov &     19.15 &   21.45 &    2.28 &         2.30 &          9.43 \\
         Periscope &     19.15 &   23.45 &    9.60 &         4.30 &         41.34 \\
         Twitter &     19.00 &   23.45 &    2.93 &         4.45 &         25.95 \\
         WhatsApp &     19.00 &   21.45 &    1.00 &         2.45 &          4.68 \\\hline
\end{tabular}
\caption{Non-normalized features values for applications spiking in Paris on the day of fire.}
\label{table:features Paris day of}
\end{table}
\newpage

\begin{table}[h!]
\centering
\begin{tabular}{|c c c c c c c|} 
 \hline
                     City & App &  $s_1^{start}$ &  $s_1^{end}$ &  $s_1^{max}$ &  $s_1^{duration}[hrs.mins]$ &  $s_1^{aggregate}$ \\
 \hline
          Marseille & Facebook Messenger &     19.30 &   22.30 &    0.98 &         3.00 &          3.28 \\
          & Periscope &     19.15 &   20.45 &    1.95 &         1.30 &          4.83 \\
               & Twitter &     19.15 &   23.45 &    0.99 &         4.30 &          7.76 \\
              \hline
            Strasbourg & Apple Video &     19.45 &   20.30 &    0.36 &         0.45 &          0.53 \\
          & DailyMotion &     19.45 &   21.00 &    0.47 &         1.15 &          0.93 \\
          & Facebook &     20.00 &   21.00 &    0.09 &         1.00 &          0.17 \\
          &  Facebook Messenger &     19.15 &   22.00 &    2.04 &         2.45 &          6.38 \\
           & Molotov &     19.45 &   20.30 &    0.63 &         0.45 &          1.26 \\
            & Periscope &     19.15 &   20.15 &    1.97 &         1.00 &          4.74 \\
            & Twitter &     19.15 &   23.45 &    0.92 &         4.30 &          4.35 \\
            \hline
        
            Lyon & Apple Video &     19.45 &   20.45 &    1.04 &         1.00 &          2.49 \\
         &Facebook Messenger &     19.30 &   21.45 &    1.77 &         2.15 &          5.33 \\
                 & Instagram &     22.00 &   23.45 &    0.03 &         1.45 &         -0.06 \\
                 &Molotov &     19.30 &   21.00 &    1.01 &         1.30 &          2.75 \\
                 & Periscope &     19.15 &   21.00 &    3.67 &         1.45 &          9.80 \\
                 & Twitter &     19.15 &   23.45 &    1.52 &         4.30 &         10.89 \\
                 \hline
       Montpellier& Apple Video &     19.45 &   21.00 &    0.70 &         1.15 &          2.27 \\
       & DailyMotion &     19.45 &   23.45 &    1.18 &         4.00 &          6.01 \\
         &  Facebook &     19.45 &   23.45 &    0.13 &         4.00 &          0.95 \\
         & Facebook Messenger &     19.30 &   22.30 &    1.81 &         3.00 &          5.14 \\
         & Instagram &     20.15 &   23.45 &    0.23 &         3.30 &          1.58 \\
           & Molotov &     20.00 &   21.00 &    0.40 &         1.00 &          0.66 \\
           & Periscope &     19.15 &   20.30 &    2.91 &         1.15 &          8.22 \\
          & Twitter &     19.15 &   23.45 &    1.32 &         4.30 &          9.29 \\
          & WhatsApp &     19.45 &   20.30 &    0.26 &         0.45 &          0.48 \\

           \hline

              Rennes & Facebook Messenger &     19.30 &   22.30 &    1.88 &         3.00 &          5.17 \\
              & Periscope &     19.15 &   20.15 &    1.23 &         1.00 &          2.12 \\
                 &Twitter &     19.15 &   21.15 &    0.61 &         2.00 &          2.16 \\ \hline
\end{tabular}
\caption{Non-normalized feature values for applications spiking in other cities on the day of fire.}
\label{table:features other cities day of}
\end{table}

\newpage

\subsubsection{Features, day after fire}
\begin{figure}[h!]
    \centering
    \includegraphics[width=\textwidth]{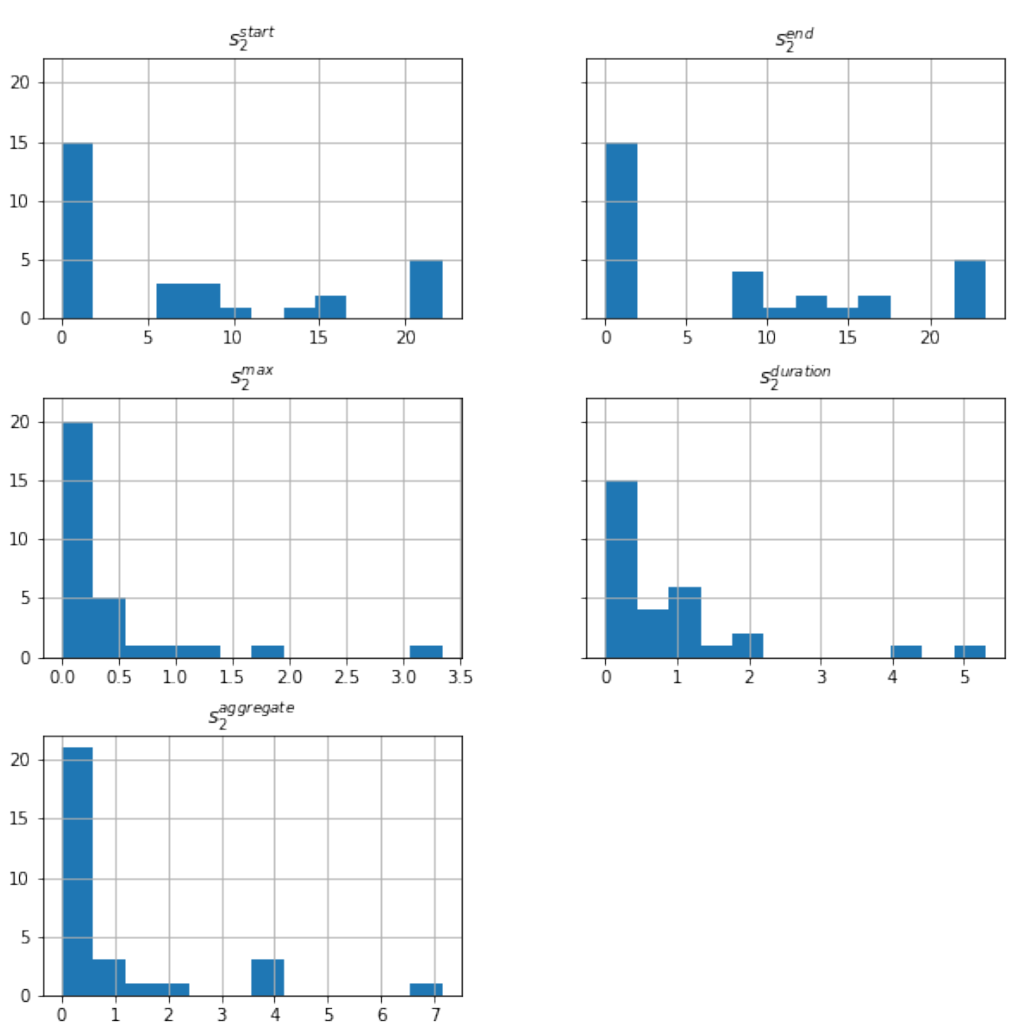}
    \caption{\textbf{Distributions of feature values}, The distribution of the values of each feature for all other cities on the day \emph{after} the fire is shown.}
    \label{fig:Dist_feat_other_day_after}
\end{figure}

A full description of the non-normalized values for the features for the day of the after fire of Notre Dame is given in Tables \ref{table:features Paris day after} and \ref{table:features other cities day after}.  Table \ref{table:features Paris day after} shows the non-normalized values for features of applications in the city of Paris and Table \ref{table:features other cities day after} shows the values for the features of spiking applications in other cities. The distribution of the values of each feature across all applications is other cities on the day after fire is shown in Figure \ref{fig:Dist_feat_other_day_after}.

\begin{table}[h!]
\centering
\begin{tabular}{|c c c c c c|} 
 \hline
                     App &  $s_2^{start}$ &  $s_2^{end}$ &  $s_2^{max}$ &  $s_2^{duration}[hrs.mins]$ &  $s_2^{aggregate}$ \\
 \hline
      Apple Video &      7.00 &   20.45 &    0.57 &        13.45 &         10.60 \\
      Apple iCloud &      7.15 &    8.00 &    0.07 &         0.45 &          0.10 \\
        DailyMotion &     15.45 &   17.15 &    1.01 &         1.30 &          4.24 \\
          Facebook &      9.00 &   11.45 &    0.03 &         2.45 &          0.18 \\
        Facebook Live &      0.00 &    0.00 &    0.00 &         0.00 &          0.00 \\
Facebook Messenger &      0.00 &    0.00 &    0.00 &         0.00 &          0.00 \\
       Instagram &      8.15 &    9.45 &    0.04 &         1.30 &          0.11 \\
        Molotov &      7.00 &   16.15 &    1.42 &         9.15 &         14.67 \\
         Periscope &      7.00 &    9.00 &    1.72 &         2.00 &          8.89 \\
         Twitter &      7.00 &   23.45 &    0.33 &        16.45 &          7.88 \\
         WhatsApp &     17.30 &   20.45 &    0.15 &         3.15 &          1.04 \\ \hline
\end{tabular}
\caption{Non-normalized features values for applications spiking in Paris on the day \emph{after} the fire.}
\label{table:features Paris day after}
\end{table}

\begin{table}[h!]
\centering
\begin{tabular}{|c c c c c c c|} 
 \hline
                     City & App &  $s_2^{start}$ &  $s_2^{end}$ &  $s_2^{max}$ &  $s_2^{duration}[hrs.mins]$ &  $s_2^{aggregate}$ \\
 \hline
          Marseille & Facebook Messenger &      9.30 &   10.15 &    0.24 &         0.45 &   0.45 \\
          &Periscope &     22.15 &   23.00 &    0.45 &         0.45 &          1.17 \\
              &Twitter &      0.00 &    0.00 &    0.00 &         0.00 &          0.00 \\
              \hline
              
            Strasbourg &Apple Video &      0.00 &    0.15 &    0.12 &         0.15 &          0.12 \\
          &DailyMotion &     16.00 &   17.00 &    0.44 &         1.00 &          1.56 \\
          &Facebook &      0.00 &    0.00 &    0.00 &         0.00 &          0.00 \\
  & Facebook Messenger &      0.00 &    0.00 &    0.00 &         0.00 &          0.00 \\
          &Molotov &      0.00 &    0.30 &    0.25 &         0.30 &          0.47 \\
            &Periscope &     21.45 &   23.00 &    1.71 &         1.15 &          3.61 \\
            & Twitter &      0.00 &    0.00 &    0.00 &         0.00 &          0.00 \\
\hline

               Lyon  &Apple Video &      7.30 &    8.30 &    0.16 &         1.00 &          0.30 \\
       & Facebook Messenger &      0.00 &    0.00 &    0.00 &         0.00 &          0.00 \\
                 &Instagram &      0.00 &    0.00 &    0.00 &         0.00 &          0.00 \\
                 & Molotov &      8.00 &   12.15 &    0.92 &         4.15 &          3.99 \\
                 &Periscope &      7.15 &    8.45 &    0.36 &         1.30 &          0.57 \\
                 & Twitter &      0.00 &    0.00 &    0.00 &         0.00 &          0.00 \\

                 \hline
       Montpellier &Apple Video &      7.45 &    8.30 &    0.26 &         0.45 &          0.35 \\
       & DailyMotion &     16.15 &   17.00 &    0.50 &         0.45 &          0.63 \\
           &Facebook &     21.45 &   23.45 &    0.10 &         2.00 &          0.19 \\
           & Facebook Messenger &      0.00 &    0.00 &    0.00 &         0.00 &          0.00 \\
           &Instagram &      0.00 &    0.00 &    0.00 &         0.00 &          0.00 \\
            &Molotov &      7.00 &    8.45 &    1.27 &         1.45 &          4.05 \\
            & Periscope &     14.15 &   15.30 &    0.40 &         1.15 &          0.54 \\
           &Twitter &      8.00 &   13.30 &    0.24 &         5.30 &          1.17 \\
           &WhatsApp &      0.00 &    0.00 &    0.00 &         0.00 &          0.00 \\
\hline

             Rennes &Facebook Messenger &      0.00 &    0.00 &    0.00 &         0.00 &   0.00 \\
             & Periscope &     21.30 &   23.00 &    3.34 &         1.30 &          7.15 \\
                 &Twitter &      0.00 &    0.00 &    0.00 &         0.00 &          0.00 \\ \hline
\end{tabular}
\caption{Non-normalized feature values for applications spiking in other cities on the day \emph{after} the fire.}
\label{table:features other cities day after}
\end{table}

\subsection{Joint Two-Day clusters}

\begin{figure}[h!]
    \centering
    \makebox[\textwidth][c]{\includegraphics[width=1.2\textwidth]{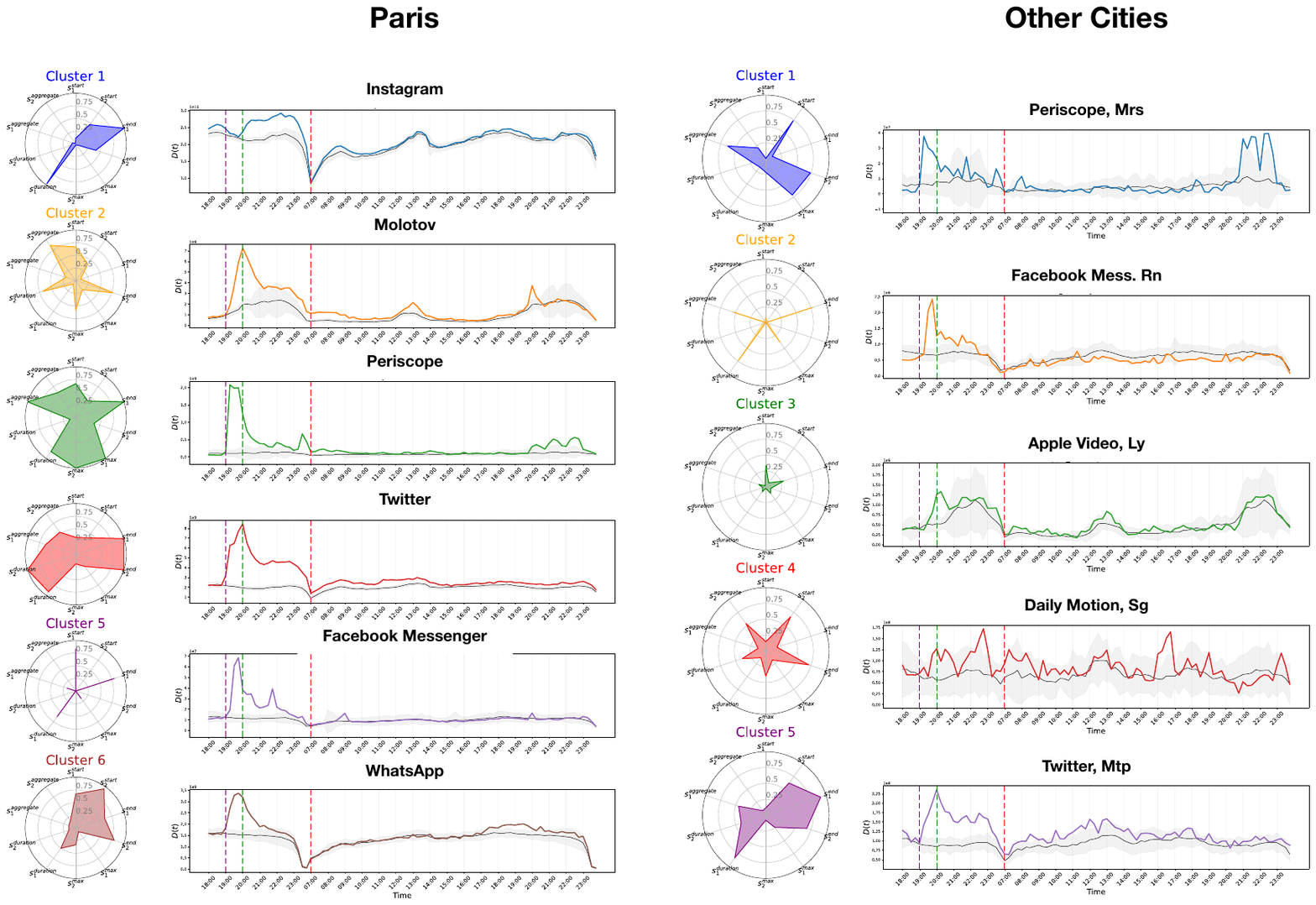}}
   \caption{\textbf{a)} For the city of Paris, the radar plots of clusters of applications are given on the left, showing the value of different features assigned to each cluster. On the right of each radar plot is a time series for an application that is representative of the applications in that cluster. Note, the information in the radar plots for each cluster is normalized within the city of Paris. The applications in each cluster are given in Table \ref{table:joint_paris}. \textbf{b)} For all other cities being considered, the radar plots of clusters of applications are given on the left, with a representative time series of an of an application in the cluster shown on the right. Note, the information in the radar plots for each cluster is normalized for all cities excluding Paris. The applications in each cluster are given in Table \ref{table:joint_other}.}
    \label{fig: joint clusters}
\end{figure}

We present the joint two-day clustering of application behavior in response to the fire of Notre-Dame. In contrast to the analysis presented in the main text, which classifies short and long term behavior of applications as separate behavioral patterns, this section considers application traffic response over an extended period of time. We define two spikes in abnormal traffic volume for an consistent with our previous analysis (see Methods for further details), recovering a 10 dimensional features vector for each spiking application. This 10 dimensional vector is composed of the union of the 5 dimensional features vectors for an application across the day of and day after the fire. We perform a k-means clustering on these features vectors after normalization, computing clustering for applications in Paris and in other cities separately, recovering 6 clusters for Paris and 5 clusters for other cities (see Figure \ref{fig: joint clusters}). 

\begin{table}[h!]
\centering
\begin{tabular}{||c c c c c c||} 
 \hline
 Cluster 1 & Cluster 2 & Cluster 3 & Cluster 4 & Cluster 5 & Cluster 6\\ [0.5ex] 
 \hline\hline
 Apple iCloud & Apple Video & Periscope & Twitter & Facebook Live & DailyMotion\\ 
 Facebook & Molotov &  &  & Facebook Messenger & WhatsApp \\
 Instagram &  &  &  & & \\[1ex] 
 \hline
\end{tabular}
\caption{Joint two-day clusters for Paris}
\label{table:joint_paris}
\end{table}

\begin{table}[h!]
\centering
\begin{tabular}{||c@{\hskip 0.1in}c@{\hskip 0.1in}c@{\hskip 0.1in}c@{}c||} 
 \hline
 Cluster 1 & Cluster 2 & Cluster 3 & Cluster 4 & Cluster 5\\ [0.5ex] 
 \hline\hline
 Periscope Mrs & Facebook Messenger Sg & Apple Video Sg & DailyMotion Sg & DailyMotion Mtp\\
 Periscope Sg & Facebook Messenger Ly & Apple Video Ly & Molotov Ly & Facebook Mtp\\
 Periscope Ly & Facebook Messenger Mtp & Apple Video Mtp &  Molotov Mtp & Facebook Messenger Mrs\\
 Periscope Mtp & Facebook Messenger Rn & Facebook Sg & Periscope Rn & Twitter Mtp\\
  & Instagram Rn & Instagram  Ly &  & \\
  & Twitter Mrs  & Molotov Sg &  & \\
  & Twitter Sg  & Twitter Rn &  & \\
  & Twitter Ly  & Whatsapp Mtp  &  & \\[1ex] 
 \hline
\end{tabular}
\caption{Joint two-day clusters for other cities, Marseille (Mrs), Lyon (Ly), Montpellier (Mtp), Rennes (Rn) and Strasbourg (Sg)}
\label{table:joint_other}
\end{table}

In the joint two-day clusters, the short and long term behavior of applications are considered a single pattern. From the joint clusters, we find similar trends to those seen in the main text. For example, for Paris, Twitter and Periscope both form an outlier cluster with a single defined behavior, as is done on the day of the fire. Likewise, for other cities, we have a clustering consisting of only the app Periscope. Interestingly, for other cities the applications Twitter and Facebook Messenger, which have distinct behaviors when considered through the perspective of short and long term behaviors, cluster together when considering the joint clustering. This is largely due to both of their abnormal traffic activity being largely confined to the day of the fire.

\end{document}